\documentclass[a4paper,11pt]{article}

\usepackage{jheppub}

\usepackage{lineno}
\usepackage{amssymb,amsmath,amsthm}
\usepackage[shortlabels]{enumitem}
\usepackage{braket}
\usepackage{comment}
\usepackage{subcaption}
\usepackage{graphicx, xcolor, varwidth}
\usepackage{color}
\usepackage{csquotes} 
\MakeOuterQuote{"}    
\usepackage{mathtools}
\usepackage{tikz}
\usepackage{commath}
\usepackage{adjustbox}

\def\O{\mathcal{O}}

\def\R{\mathbb{R}}

\def\p{\partial}  

\def\<{\langle}
\def\>{\rangle}

\newcommand{\Seff}{S_{\mathrm{eff}}}
\newcommand{\xv}{\bd{x}}
\newcommand{\xvp}{\bd{x}{'}}
\newcommand{\xvpp}{\bd{x}{''}}
\newcommand{\intx}{\int d^{2}\xv\,}
\newcommand{\intxx}{\int d^{2}\xv\,d^{2}\xvp\,}

\newcommand{\intxpp}{\int d^{2}\xv''}
\newcommand{\hq}{\hat{q}}
\newcommand{\hL}{\hat{\Lambda}}

\usepackage{tikz}
\newcommand*{\boxcolor}{black}
\makeatletter
\renewcommand{\boxed}[1]{\textcolor{\boxcolor}{%
\tikz[baseline={([yshift=-1ex]current bounding box.center)}] \node [rectangle, minimum width=1ex,rounded corners,draw] {\normalcolor\m@th$\displaystyle#1$};}}
 \makeatother
 
\def\Tr{\text{Tr}}
\newcommand{\bd}{\boldsymbol}

\usepackage{fancybox} 
\title{\boldmath The two-dimensional disordered $O(N)$ sigma model}

\author[a]{Martí Berenguer,}
\author[b]{Felix M. Haehl,}
\author[c]{and Elisa Tabor}

\affiliation[a]{Departamento de Física de Partículas, Universidade de Santiago de Compostela and Instituto
Galego de Física de Altas Enerxías (IGFAE). E-15782 Santiago de Compostela, Spain}
\affiliation[b]{School of Mathematical Sciences \& STAG Research Centre, \\
University of Southampton, SO17 1BJ, Southampton, United Kingdom}
\affiliation[c]{Leinweber Institute for Theoretical Physics, Stanford University, Stanford, CA 94305, USA}

\emailAdd{marti.berenguer.mimo@usc.es}
\emailAdd{f.m.haehl@soton.ac.uk}
\emailAdd{etabor@stanford.edu}

\abstract{
We introduce a two-dimensional $O(N)$ nonlinear sigma model with random Gaussian $p$-body interactions. The model combines the structure of a two-dimensional bosonic SYK-type quantum field theory with the stabilizing spherical constraint of the nonlinear sigma model. At large $N$ we derive the Schwinger-Dyson equations on a torus and analyze the solutions using both analytical approximations and numerical methods. We find a phase diagram qualitatively similar to that of the one-dimensional quantum spherical $p$-spin model, including a low-temperature transition to a spin glass phase. This phase is characterized by a finite Edwards-Anderson order parameter, while the dynamical part of the two-point function displays an approximate scaling regime. These results provide a tractable setting for studying approximate conformal behavior and glassy physics in a two-dimensional relativistic field theory with disorder.}

\usepackage[T1]{fontenc}
\usepackage{lmodern}
\begin{document}
\maketitle

\setcounter{tocdepth}{2}

\section{Introduction}

Over the past decade, the SYK model \cite{Sachdev:1992fk,Kitaev:2015,Maldacena:2016hyu,Kitaev:2017awl} has played a crucial role in furthering our understanding of strange metals and low-dimensional holography (e.g., \cite{Sachdev:2015efa,maldacena:2016,Saad:2019lba}). Crucially, the model is microscopically defined by an ensemble average over disorder and it exhibits a near-conformal strong-coupling limit, where its thermodynamics is dominated by a soft mode associated with the breaking of conformal symmetry.

While the holographic description of the Schwarzian dynamics is particularly clear, it is substantially more challenging to export these lessons to higher dimensions. It would be of great interest to identify a two-dimensional conformal field theory (CFT$_2$) that describes aspects of semiclassical gravity in asymptotically AdS${}_3$ spacetimes in an analogous fashion. To this end, various two-dimensional generalizations of the SYK model have been proposed, e.g., \cite{Berkooz:2016cvq,Turiaci:2017zwd,Berkooz:2017efq,Murugan:2017eto,Liu:2018jhs,Biggs:2026zir}. For example, the Murugan-Stanford-Witten (MSW) models \cite{Murugan:2017eto} are disordered two-dimensional field theories with emergent conformal (Virasoro) symmetry at strong coupling. While the symmetry-breaking mechanism is not identical to the SYK model (in particular, there is no need to break the conformal symmetry explicitly and the associated Goldstone mode is not parametrically enhanced), the formalism nevertheless exposes similarities. For example, the models admit a mean-field description with Schwinger-Dyson equations that are dominated by a solvable class of ``melon diagrams''.\footnote{Further, the two-dimensional analog of the Schwarzian soft mode action is expected to provide an effective description of the stress tensor sector in this model \cite{toappearFelix} (see also \cite{Altland:2025qqw}), thus connecting it to the mean-field description of AdS$_3$ boundary gravitons \cite{Cotler:2018zff}.} A crucial question that motivates our work is whether the conformal phase in the MSW models or similar disordered field theories is in fact physically realized. While this is expected to be the case in supersymmetric models, the present work addresses this question in the case of bosonic microscopic degrees of freedom. If a conformal phase can be realized, then these models would provide a rare and valuable instance of explicit chaotic large-$N$ CFT$_2$ with holographic characteristics.

To see why this question is nontrivial to answer, let us emphasize what may obstruct a realization of the emergent conformal symmetry in the strong-coupling limit: recall that disordered models with bosonic microscopic degrees of freedom tend to undergo a phase transition to a spin glass at low temperatures, as has been known for a long time (see, e.g., \cite{Castellani_2005,Chowdhury:2021qpy} for overviews). In order to exhibit the SYK-like physics of non-Fermi liquids (such as finite ground state entropy, strong chaos etc.), it is important to avoid a spin-glass phase. On the other hand, spin glasses are interesting in their own right and a holographic description of spin glass order is also an important open problem; for interesting suggestions and progress see \cite{Anninos:2011vn,Anninos:2012gk,Anninos:2013mfa,Anninos:2016szt,Engelhardt:2020qpv,Anous:2021eqj}. Indeed, the prospect of identifying spin glass models with a tractable holographic description serves as a second motivation for this work: the authors are not aware of detailed studies of glassy physics in genuine quantum field theories with holographic characteristics.

Motivated by these observations, in this work we study a two-dimensional generalization of the $p$-spin spherical model \cite{Cugliandolo_2000,Cugliandolo_2001,Baldwin:2019dki,Anous:2021eqj} (see also \cite{Bera:2021lnh,Winer:2022ciz,Correale_2023}), which we refer to as the {\it disordered $O(N)$ sigma-model}. The model consists of $N\gg 1$ bosonic rotor degrees of freedom $\sigma_i$ subject to disordered interactions and a stabilizing spherical constraint. That is, the Euclidean action is of the following form:
\begin{equation}
\label{eq:modelDef0}
 S =  \int_{{\cal M}} d^2 \bd{x} \, \left[ \frac{1}{2}\, (\partial_\mu \sigma_i)^2 + \frac{1}{2} \, \lambda \left( \sigma_i^2 - \frac{N}{g_0^2}\right) + \sum_{i_1 < \cdots < i_p} J_{i_1 ,\ldots , i_p} \, \sigma_{i_1} \cdots \sigma_{i_p}\right] \,.
\end{equation}

The spherical constraint (second term) imposes that the vector of bosonic spins must lie on a sphere with fixed radius $\sqrt{N/g_0^2}$, i.e., the model is a sigma model with high-dimensional spherical target space. The couplings $J_{i_1 ,\ldots , i_p}$ are averaged over with a Gaussian probability distribution whose variance defines an average coupling strength $J$. At strong coupling, the one-dimensional version of the model naively exhibits a conformal phase with emergent time reparametrization symmetry \cite{Cugliandolo_2000,Anous:2021eqj}. However, this phase is not physically realized: depending on the size of the spherical target space, either a different (non-conformal) solution branch is thermodynamically preferred, or the model undergoes a phase transition into a spin glass phase. We will study whether similar behavior occurs in two dimensions. To this end, it will be most natural to study the model on a torus manifold ${\cal M} = \mathbb{T}^2$.

Note that the model \eqref{eq:modelDef0} is closely related to the bosonic MSW model \cite{Murugan:2017eto}: it only differs by the addition of the Lagrange multiplier term (spherical constraint), which plays the same role as a (running) mass term. Indeed, in the large-$N$ limit, we can ignore fluctuations of $\lambda$ and consider it constant, $\langle \lambda(\xv) \rangle = \lambda$.
The bosonic MSW model is not well-defined as it exhibits an instability due to runaway directions in a bosonic disordered potential. The spherical constraint cures this problem; we expect (and will indeed show) that \eqref{eq:modelDef0} defines a consistent and stable theory.\footnote{A different cure is provided by a supersymmetric extension of the model \cite{Murugan:2017eto}. We understand that this is being investigated elsewhere \cite{BerkoozToAppear}.}

\begin{figure}
    \centering
        \includegraphics[width=\textwidth]{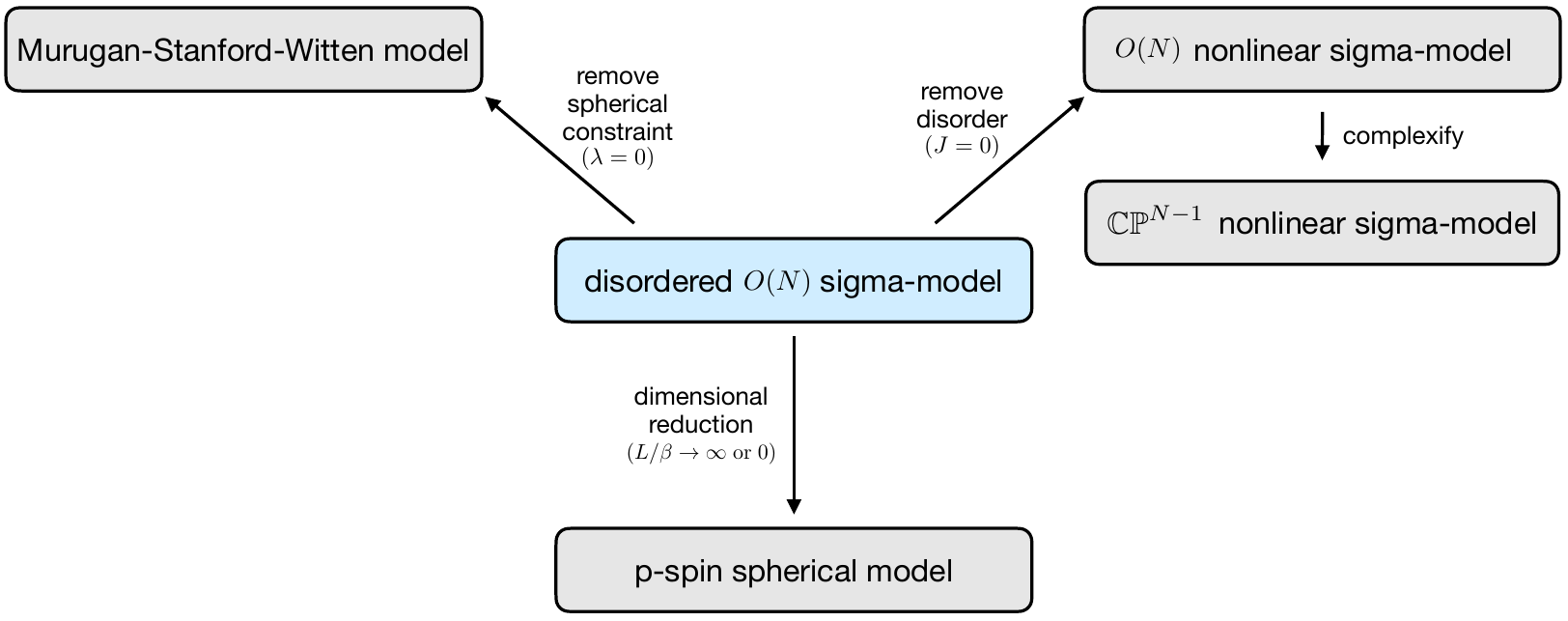}
    \caption{The relation between the disordered $O(N)$ sigma model studied in this paper and various models in the literature.}
    \label{fig:models}
\end{figure}

Alternatively, our model can also be thought of as a generalization of the well-known $O(N)$ nonlinear sigma model \cite{Polyakov:1975rr,Brezin:1976ap,Coleman:1985rnk} coupled through spacetime-independent disorder.\footnote{This should be contrasted against spacetime-dependent disorder, for example as an RG perturbation of a UV conformal field theory, cf.\ \cite{Imry:1975zz,Aharony:2015aea}.} In the case of the $O(N)$ model, the fundamental degrees of freedom are also bosonic rotors $\sigma_i$ subject to a spherical constraint. We make extensive use of formalism developed in disordered quantum mechanics (such as the SYK model and the $p$-spin spherical model) as well as the $O(N)$ nonlinear sigma model. We also draw inspiration from the $\mathbb{CP}^{N-1}$ nonlinear sigma model, which differs only by the fact that the rotors $\sigma_i$ are complex: the spherical constraint is then subject to a $U(1)$ gauge redundancy and the target space is $\mathbb{CP}^{N-1}$ instead of $S^{N-1}$. See \cite{DAdda:1978vbw,Witten:1978bc} for original work on the $\mathbb{CP}^{N-1}$ model and \cite{Monin:2015xwa,Bolognesi:2016zjp,Betti:2017zcm,Bolognesi:2019rwq} for related discussions.
In Fig.\ \ref{fig:models}, we summarize the various models and relations between them.\footnote{Our model can also be viewed in the context of tensor models \cite{Gurau:2009tw,Gurau:2011xp,Witten:2016iux,Klebanov:2016xxf}: these models do not integrate over disorder but nevertheless exhibit large-$N$ dynamics dominated by melon diagrams, as in the SYK model (see also \cite{Biggs:2026zir}).}

The paper is organized as follows. In Section \ref{sec:review}, we begin by reviewing the large-$N$ solution of the $O(N)$ nonlinear sigma model, focusing in particular on the renormalization of the spherical constraint in two dimensions. In Section \ref{sec:disordered}, we add $p$-body SYK-like interactions to this model and derive its large-$N$ Schwinger-Dyson equations using a one-step replica-symmetry-breaking ansatz. In Section \ref{sec:staticanalysis}, we perform a simplified analysis of the equations of motion usually referred to as the "static" approximation. In Section \ref{sec:numerical}, we solve the full Schwinger-Dyson equations numerically and determine the phase diagram and thermodynamics. In Appendix \ref{app:EoM}, we include more details on the derivation of the equations of motion of the two-dimensional disordered $O(N)$ sigma model. In Appendix \ref{app:FreeEnReg}, we explain the regularization of the on-shell free energy. In Appendix \ref{app:numericalmethod}, we give details on the numerical solution of the Schwinger-Dyson equations. In Appendix \ref{app:dim_red}, we show how the model reduces to the one-dimensional quantum $p$-spin model \cite{Anous:2021eqj} by considering an appropriate dimensional reduction.

\section{Review of the \texorpdfstring{$\bd{O(N)}$}{O(N)} nonlinear sigma model} \label{sec:review}

Much of the technical and numerical analysis in later sections can be illustrated in the well-known $O(N)$ nonlinear sigma model. The latter corresponds to our model in the limit of zero disordered coupling (see Fig.\ \ref{fig:models}). We sometimes refer to this as the "free limit", with the understanding that we still keep the interactions due to the spherical constraint intact.

The Euclidean action of the $(1+1)$-dimensional $O(N)$ nonlinear sigma model is \cite{Polyakov:1975rr,Brezin:1976ap,Coleman:1985rnk}
\begin{equation} \label{eq: free Lagrangian}
    S = \int_{{\cal M}} d^2 \bd{x} \, \left[ \frac12 (\p_\mu\sigma_i)^2 + \frac12\,\lambda\left(\sigma_i^2 -\frac{N}{g_0^2}\right) \right]\,,
\end{equation}
where $\sigma_i$ are $N$ bosonic fields, $g_0$ is the bare coupling constant, and $\lambda(\xv)$ is a Lagrange multiplier enforcing the spherical constraint
\begin{equation} \label{eq:sph_constraint}
    \sum_{i=1}^N \sigma_i^2 = \frac{N}{g_0^2} \,.
\end{equation}
The fields therefore live on an $(N-1)$-dimensional sphere $S^{N-1} = O(N) / O(N-1)$ of radius $\sqrt{N}/g_0$.

The partition function of the model is
\begin{equation}\label{eq:O(N)partitionfunction}
    Z = \int \mathcal{D}\sigma_i\int_{-i\infty}^{i\infty}\mathcal{D}\lambda\,e^{-S[\sigma,\lambda]}\,,
\end{equation}
where the integration contour for $\lambda(\xv)$ is chosen along the imaginary axis in order to make the integration finite.\footnote{This convention is equivalent to the more familiar representation of the constraint through a delta function, in which the auxiliary field is integrated along the real axis and appears in the action as $i\lambda$. Throughout this work we adopt the convention \eqref{eq:O(N)partitionfunction}, for which the saddle-point value of $\lambda$ directly coincides with the dynamically generated mass squared (cf. Eq. \eqref{eq:Lambda}).}

\paragraph{$\bd{O(N)}$ model on the Euclidean plane.}
We first consider the theory \eqref{eq: free Lagrangian} on ${\cal M}=\R^2$. In the large-$N$ limit we set $\lambda(\xv)=\lambda$. Integrating out the $N$ bosonic fields yields the effective action
\begin{equation}\label{eq:SeffONmodelplane}
    Z=\int\mathcal{D}\lambda \, e^{-\Seff[\lambda]}\,,\qquad \frac{\Seff[\lambda]}{N\,V_2}=-\frac{\lambda}{2g_0^2}+ \frac{1}{2}\int \frac{d^2\mathbf{k}}{(2\pi)^2}\log({\mathbf{k}^2 + \lambda})\,,
\end{equation}
where $\bd{k}^2=\omega^2+p^2$ and $V_2$ denotes the (infinite) two-dimensional volume.

The saddle-point equation takes the form
\begin{equation} \label{eq:bareplanar}
    \frac{1}{g_0^2} = \int \frac{d^2\bd{k}}{(2\pi)^2}\,\hq(\bd{k}) \,,\qquad \hq(\bd{k})=\frac{1}{\bd{k}^2+\lambda}\,.
\end{equation}
This equation is the large-$N$ spherical constraint. The momentum integral is logarithmically divergent. Introducing an ultraviolet cutoff $|\bd{k}|<\Lambda_{\rm UV}$,
\begin{equation}\label{eq:cutoffconstraint}
    \frac{1}{g_0^2} = \frac{1}{4\pi}\log\frac{\Lambda_{\rm UV}^2}{\lambda}+\mathcal{O}(\lambda/\Lambda_{\rm UV}^{2})\,.
\end{equation}
The divergence can be absorbed into a renormalized coupling defined at an arbitrary scale $\mu$,
\begin{equation}\label{eq:running}
\frac{1}{g(\mu)^2}= \frac{1}{g_0^2} - \frac{1}{4\pi}\log\frac{\Lambda_{\rm UV}^2}{\mu^2}\,.
\end{equation}
Substituting \eqref{eq:running} into \eqref{eq:cutoffconstraint} gives
\begin{equation}
    \frac{1}{g(\mu)^2} = \frac{1}{4\pi}\log\frac{\mu^2}{\lambda}\,.
    \label{eq:renconstraint}
\end{equation}
The theory is asymptotically free: the coupling
vanishes logarithmically as $\mu\to\infty$, and grows strong in the infrared. 
The solution of~\eqref{eq:renconstraint} for $\lambda$ is
\begin{equation}\label{eq:Lambda}
    \lambda = \mu^2 e^{-4\pi/g(\mu)^2}\equiv \Lambda^2\,.
\end{equation}
Through this relation, the dimensionless coupling $g$ has been traded for the dynamically generated scale $\Lambda$, a standard example of dimensional transmutation \cite{Coleman:1985rnk, Bolognesi:2016zjp}. Note that the solution \eqref{eq:Lambda} for $\lambda$ is invariant under the running of the coupling $g$ with $\mu$, and all dependence on the UV cutoff $\Lambda_{\rm UV}$ has disappeared. 

At the saddle point, the propagator becomes
\begin{equation}
    \hat{q}(\bd{k})=\frac{1}{\bd{k}^2+\Lambda^2} \,,
    \label{eq:qplanar}
\end{equation}
which takes the form of a massive propagator with mass $\Lambda$. Dimensional transmutation therefore has a direct physical consequence: the fields $\sigma_i$, which are massless at the level of the classical action, acquire a dynamical mass $\Lambda$ through the strong-coupling dynamics of the infrared.

The corresponding position-space correlator is
\begin{equation}
    q(\bd{x})=\int \frac{d^2\bd{k}}{(2\pi)^2}\frac{e^{-i\bd{k}\cdot \bd{x}}}{\bd{k}^2+\Lambda^2}=\frac{1}{2\pi}K_0\left(\Lambda|\bd{x}|\right)\,,
    \label{eq:qposplanar}
\end{equation}
where $K_0$ is a modified Bessel function of the second kind. This correlator has the usual short-distance divergence present in two-dimensional QFTs:
\begin{equation}
    q(\bd{x}) = -\frac{1}{2\pi}\log(\Lambda|\bd{x}|) + \text{finite} \qquad (|\bd{x}| \rightarrow 0) \,,
\end{equation}
while at large separation,
\begin{equation}
    q(\bd{x})=\frac{1}{\sqrt{8\pi\Lambda \abs{\bd{x}}}}\,e^{-\Lambda \,\abs{\bd{x}}}+\ldots \qquad (\,\abs{\bd{x}}\to\infty) \,.
\end{equation}

It is also useful to introduce the spectral representation $\rho(k)$ as
\begin{equation}
    q(\bd{x})=\int_\Lambda^\infty \frac{dk}{2\pi}\,\rho(k)\,e^{-k\,\abs{\bd{x}}}\,,\qquad \rho(k)=\frac{1}{\sqrt{k^2-\Lambda^2}} \,,
\end{equation}
where $k\equiv|\bd{k}|$. The gap $\Lambda$ therefore controls the exponential decay of the correlator and may equivalently be identified with the inverse zero mode,
\begin{equation}
    \frac{1}{\hq(\bd{0})}=\Lambda^2\,.
\end{equation}

This gapped behavior at late times presents an obstruction to conformality. Consequently, for the correlators to approach the conformal regime of the bosonic MSW model \cite{Murugan:2017eto}, we would require the vanishing of the gap $1/\hq(\bd{0})$.

\paragraph{$\bd{O(N)}$ model on a torus geometry.}

\begin{figure}
    \centering
    \begin{subfigure}[t]{0.48\textwidth}
        \centering
        \includegraphics[width=\textwidth]{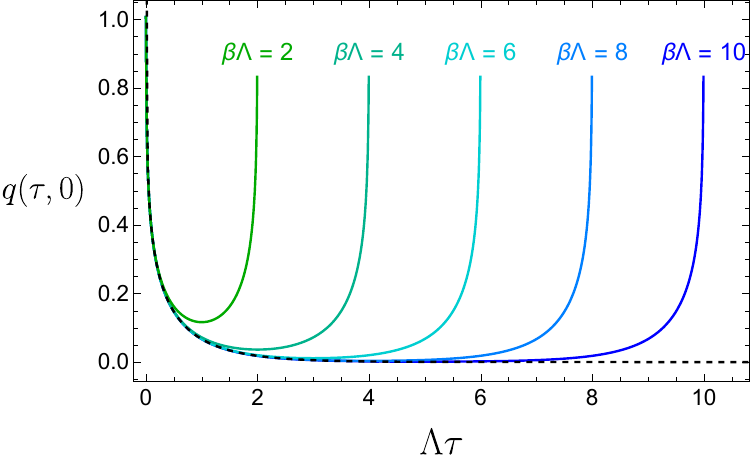}
    \end{subfigure}
    \hspace{0.02\textwidth}
    \begin{subfigure}[t]{0.48\textwidth}
        \centering
        \includegraphics[width=\textwidth]{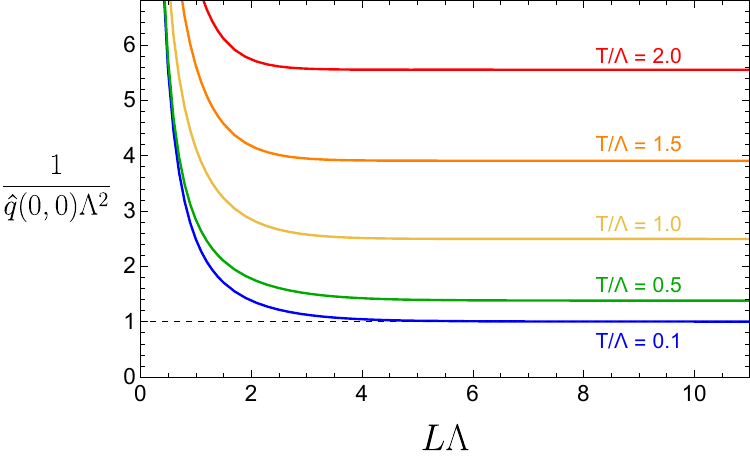}
    \end{subfigure}
    \caption{{\it Left}: Position-space correlator $q(\tau,x=0)$ approaching the planar limit, with the analytic result \eqref{eq:qposplanar} shown in dashed. {\it Right}: Inverse zero modes on the torus as a function of $L$ for different temperatures, obtained numerically using the renormalized spherical constraint \eqref{eq:sphconstrren}. Results agree with those from Pauli-Villars regularization (see Fig. 2 of \cite{Bolognesi:2019rwq}).}
    \label{fig:CPN-1}
\end{figure}

We now consider the theory \eqref{eq: free Lagrangian} on a flat torus ${\cal M}=\mathbb{T}^2$ with Euclidean time period $\beta$ and spatial period $L$. Frequencies and momenta become discrete and are given by $\omega_k=2\pi k/\beta$ and $p_l=2\pi l/L$, respectively. The effective action \eqref{eq:SeffONmodelplane} becomes
\begin{equation}
    \frac{\Seff[\lambda,\beta,L]}{N\,\beta L}=-\frac{\lambda}{2g_0^2}+ \frac{1}{2\beta L}\sum_{k,l}\log\left[\left(\frac{2\pi k}{\beta}\right)^2 +\left(\frac{2\pi l}{L}\right)^2 + \lambda\right]\,,
\end{equation}
and the saddle-point equation for $\lambda$ is now
\begin{equation}
    \frac{1}{g_0^2} = \sum_{k,l} \frac{\hat{q}(k,l)}{\beta L}\,,\qquad \hat{q}(k,l) = \frac{1}{\left(\frac{2\pi k}{\beta}\right)^2+\left(\frac{2\pi l}{L}\right)^2+ \lambda} \,,
    \label{eq:sphconstr}
\end{equation}
with $\lambda=\frac{1}{\hat q(0,0)}$. The ultraviolet divergence of \eqref{eq:sphconstr} is identical to that of the planar theory. Using the renormalization prescription \eqref{eq:running}, we obtain the renormalized spherical constraint

\begin{equation}
    \frac{1}{g(\mu)^2} =\sum_{|\bd{k}|^2<\Lambda_{UV}^2}\frac{\hq(k,l)}{\beta L}- \frac{1}{4\pi}\log\frac{\Lambda_{\rm UV}^2}{\mu^2}   \, ,
    \label{eq:sphconstrren}
\end{equation}
where $|\bd{k}|^2<\Lambda_{UV}^2$ denotes that we restrict our sum to momenta below the UV cutoff, i.e. $\omega_k^2+p_l^2<\Lambda_{UV}^2$.

In the planar limit $\beta,L\to\infty$, the discrete sum reduces to the integral in \eqref{eq:bareplanar}, and the solution for the zero mode approaches $\lambda\to\Lambda^2$. For finite $\beta$ and $L$, $\Lambda^2$ sets a lower bound on the inverse zero-mode:
\begin{equation}
    \frac{1}{\hat{q}(0,0)}\ge \Lambda^2=\lim_{\beta,L\to \infty}\; \frac{1}{\hat{q}(0,0)}\,,
    \label{eq:zmplanar}
\end{equation}
in agreement with the analysis of \cite{Bolognesi:2019rwq}, where the spherical constraint was regularized using Pauli-Villars. The resulting zero-modes and position-space correlators are shown in Fig. \ref{fig:CPN-1}, where they smoothly approach the infinite-volume solution \eqref{eq:qposplanar} as $\beta,L\to\infty$.

\section{The two-dimensional disordered \texorpdfstring{$\bd{O(N)}$}{O(N)} sigma model} \label{sec:disordered}

In this section, we add $p$-body SYK-like interactions to the $O(N)$ nonlinear sigma model. We define the two-dimensional disordered $O(N)$ sigma model as
\begin{equation}
    S = \int_{\cal M}d^2\xv\,\Bigl[\frac{1}{2}(\partial_\mu\sigma_i)^2
    + \frac{\lambda}{2}\!\left(\sigma_i^2 - \frac{N}{g_0^2}\right)
    + \sum_{i_1 < \cdots < i_p} J_{i_1\cdots i_p}\,
      \sigma_{i_1}\cdots\sigma_{i_p}
  \Bigr],
\label{eq:action}
\end{equation}
where the couplings $J_{i_1\cdots i_p}$ are independent Gaussian random variables sampled from the distribution
\begin{equation}
    P\left[J_{i_1 \cdots i_p}\right] \propto \exp\left(-\frac{N^{p-1}}{p!}\frac{J^2_{i_1\cdots i_p}}{J^2}\right)\,,
\end{equation}
and from now on we choose the space to be the flat torus ${\cal M}=\mathbb{T}^2$ with periodicities $\beta$ and $L$. Via the dimensional transmutation described in the previous section, we can interpret the Lagrange multiplier term as adding a mass counterterm to the two-dimensional bosonic MSW model of \cite{Murugan:2017eto}. A mass counterterm was indeed suggested by \cite{Murugan:2017eto} as a method of canceling divergences in numerical attempts to reach a conformal ansatz in the IR. Here, such a term arises naturally via the spherical constraint and the renormalization explained in the previous section will be crucial for our construction of a solution in the strong-coupling limit.

\subsection{Equations of motion and thermodynamic quantities}

We proceed by formulating a collective field description of the model \eqref{eq:action}.

\paragraph{Replica action.} Before proceeding any further, we introduce one more tool \cite{mezard1987spin,Monasson1995StructuralGT,MEZARD1999352,zamponi2010mean}: due to the possibility of a spin glass transition, it turns out to be useful to introduce a `chemical potential' in the state space of the system: we clone the system $m$ times (labeled by $a,b=1,\ldots,m$) and introduce a small bias term in the action, which couples the degrees of freedom in different copies in an attractive fashion:
\begin{equation}
\label{eq:Scloned}
\begin{aligned}
    S_m &= \sum_{a=1}^m \intx\left[\frac{1}{2}(\partial_\mu\sigma^a_i)^2+\frac{\lambda^a}{2}\left( (\sigma_i^a)^2-\frac{N}{g_0^2}\right) +\sum_{i_1 < \cdots < i_p}J_{i_1 \cdots i_p}\sigma^a_{i_1}\cdots \sigma^a_{i_p}\right] \\
    &\quad + \epsilon\sum_{a,b=1}^m \int d^2\bd{x} \; \sigma^a_i\sigma^b_i \,.
\end{aligned}
\end{equation}
Note that all $m$ copies have the same couplings $J_{i_1\cdots i_p}$. The last term favors for all the $m$ copies of $\sigma_i$ to be in the same thermodynamic state. A sharp diagnostic for spin glass order is whether or not correlations between the clones remain after sending $\epsilon\rightarrow 0$. We will conveniently choose the external parameter $m$ such as to achieve a dynamically stable configuration.

For ease of notation, we set $\epsilon=0$ now, but we keep the $m$-copy action.
The large-$N$ effective action for the cloned model \eqref{eq:Scloned} can then be derived using well-known path integral techniques, yielding the following averaged potential $\overline{\Phi}$:
\begin{equation}
\label{eq:PhiDef}
    \beta m \overline{\Phi} \equiv -\overline{\log Z[J_{i_1\cdots i_p}]^m} 
    =  - \lim_{n\rightarrow 0} \, \partial_n \, \overline{Z[J_{i_1\cdots i_p}]^{n\times m}} \,,
\end{equation}
where $Z[J_{i_1\cdots i_p}]$ is the partition function and $\overline{(\,\cdot\,)}$ refers to the disorder average over $J_{i_1\cdots i_p}$. In addition to the original $m$ copies, we replicate the entire system $n$ times to compute the logarithm; the second equality in \eqref{eq:PhiDef} is the standard replica trick for quenched free energies. We extend the indices $a,b=1,\ldots,mn$ to label the combined replica space. In \eqref{eq:PhiDef}, we denote the potential by $\Phi$ to suggest that it is computed in the grand-canonical ensemble where the number of clones of the system $m$ may fluctuate. 

In order to account for the possibility of spin glass order, we must perform the disorder average carefully in the replicated field space and allow for the possibility of replica off-diagonal correlations (accounting for the fact that in general $\overline{Z^{mn}} \neq \bar{Z}^{mn}$).
In order to do so, one introduces the mean-field (Hubbard-Stratonovich) variable
\begin{equation}\label{eq:defQab}
    Q_{ab}(\bd{x},\bd{x}') \equiv \frac1N \sum_{i=1}^N \sigma_i^a(\bd{x}) \sigma_i^b(\bd{x}')
\end{equation}
with replica indices $a,b=1,\ldots,mn$.

In Appendix \ref{app:EoM} we review how to integrate out $J_{i_1\cdots i_p}$ and $\sigma_i^a$ in order to obtain the large-$N$ effective action
\begin{equation}\label{eq:SefflargeN}
\begin{aligned}
    \frac{\Seff}{N}
  &= \frac{1}{2}\Tr\log\left[\frac{1}{2\pi}\delta_{ab}\,\delta^{(2)}(\xv-\xvp)\!\left(-\partial_\mu^2 + \lambda^a(\xv)\right)+\frac{i}{\pi}\Sigma_{ab}(\xv,\xvp)\right]\\
  &\quad- \frac{1}{2g_0^2}\sum_{a=1}^{nm}\intx\lambda^a(\xv)-\sum_{a,b=1}^{nm}\intxx\!\left[i\,\Sigma_{ab}(\xv,\xvp)\,Q_{ab}(\xv,\xvp)+\frac{J^2}{4}\,Q_{ab}(\xv,\xvp)^p\right]\,,
\end{aligned}
\end{equation}
where $\Sigma_{ab}(\xv,\xvp)$ is introduced as a Lagrange multiplier to enforce the definition \eqref{eq:defQab} (see Eq.\ \eqref{eq:insert1}).

It is well-known in quantum mechanical models that a replica symmetry breaking (RSB) ansatz can be made, whose consistency will be confirmed {\it a posteriori} by establishing consistent thermodynamics. As in the one-dimensional $p$-spin model, it turns out to be sufficient to make a one-step RSB ansatz of the form
\begin{equation}
\label{eq:1rsb}
	Q_{ab}(\xv,\xvp) =\begin{pmatrix} 
	q(\bd{x},\bd{x}') & u & u &   &  & &\\
	u & q(\bd{x},\bd{x}') &u &  & 0& &\cdots\\
	u & u & q(\bd{x},\bd{x}') & & & &\\
	  &  &  & q(\bd{x},\bd{x}') &  u& u &\\
	  &  0&  & u & q(\bd{x},\bd{x}') &u &\\
	  &  &  & u & u & q(\bd{x},\bd{x}') &\\
	  & \vdots & & & & & \ddots &
	\end{pmatrix}
\end{equation}
where $u$ is a dynamically determined constant, also known as Edwards-Anderson order parameter. The size of the square matrix is $mn\times mn$ and each RSB ``block'' involving $u$ and $q$ has size $m\times m$, thus representing one replica of the $m$-fold cloned system. We are interested in detecting non-trivial off-diagonal correlations between the $m$ clones through the parameter $u$. Correlations between different copies of the cloned system (block-off-diagonal entries) can be set to zero without loss of generality \cite{Anous:2021eqj}.

Using the 1-RSB ansatz, the effective action \eqref{eq:SefflargeN} becomes (in momentum space):
\begin{equation}
\begin{aligned}
   \frac{S_{\rm eff}}{nmN}&= -\frac{1}{2}\sum_{(k,l)\neq(0,0)}\log\!\frac{\hq(k,l)}{\beta L}-\frac{1}{2}\log\!\left[\frac{\hq(0,0)}{\beta L}-u\right]+\frac{1}{2m}\log\left[\frac{\frac{\hq(0,0)}{\beta L}-u}{\frac{\hq(0,0)}{\beta L}+u(m-1)}\right]\\
    &\quad+\frac{1}{2}\,\hat\lambda(0,0)\left(\sum_{k,l}\frac{\hq(k,l)}{\beta L} - \frac{1}{g_0^2}\right)+\frac{1}{2}\sum_{k,l}\!\left[\left(\frac{2\pi k}{\beta}\right)^2+\left(\frac{2\pi l}{L}\right)^2\right]\hq(k,l)\\
    &\quad-\frac{(\beta L J)^2}{4}\!\Biggl[(m-1)\,u^p+ \frac{1}{\beta L}\intx\,q(\bd{x})^p\Biggr]\,.
    \label{eq:Seff}
\end{aligned}
\end{equation}

\paragraph{Equations of motion.} The equations of motion are obtained by varying with respect to $q$, $\lambda$, and $u$. It will be convenient to define the self-energy $\Lambda(\xv,\xvp)$,
\begin{equation}
    \Lambda(\xv,\xvp)\equiv\frac{p}{2}q(\xv,\xvp)^{p-1}\,,
\end{equation}
as well as the "reduced" variables
\begin{equation}
    q_r(\xv,\xvp)\equiv q(\xv,\xvp)-u\,,\qquad \Lambda_r(\xv,\xvp) \equiv\Lambda(\xv,\xvp)-\frac{p}{2}u^{p-1}\,.
\end{equation} 
In terms of these reduced variables, the equation of motion for $\hq(k,l)$ obtained from the action \eqref{eq:Seff} adopts a very simple form (see Appendix \ref{app:EoM} for details):
\begin{equation}
    \frac{1}{\hat{q}_r(k,l)}-\frac{1}{\hat{q}_r(0,0)} = \left(\frac{2\pi k}{\beta}\right)^2+\left(\frac{2\pi l}{L}\right)^2-J^2\left(\hat{\Lambda}_r(k,l)-\hat{\Lambda}_r(0,0)\right)\,,
    \label{eq:EoMqr}
\end{equation}
while the zero modes, Lagrange multiplier, and $(u,m)$ are constrained by the following algebraic relations:
\begin{align}
    \frac{\hat{\lambda}(0,0)}{\beta L} &= \frac{1}{\hat{q}_r(0,0)} + J^2 \hat{\Lambda}_r(0,0)\,, \label{eq:zSol}\\
    0&=\frac{m-1}{2}\left[\frac{(\beta LJ)^2}{2}p\,u^{p-1}-\frac{u}{\frac{\hat{q}_r(0,0)}{\beta L}\left(\frac{\hat{q}_r(0,0)}{\beta L}+mu\right)}\right]\,. \label{eq:usol}
\end{align}

The equation obtained from varying with respect to $\lambda$ again gives the spherical constraint. Since the introduction of the random couplings does not change the ultraviolet structure of the momentum correlator, the same renormalization performed for the $O(N)$ non-linear sigma model applies in this case:
\begin{equation}
    \frac{1}{g(\mu)^2} =\sum_{|\bd{k}|^2<\Lambda_{UV}^2}\frac{\hq(k,l)}{\beta L}- \frac{1}{4\pi}\log\frac{\Lambda_{\rm UV}^2}{\mu^2}\, , \label{eq:renormalized constraint}
\end{equation}
with $\hq(k,l)$ now satisfying Eq. \eqref{eq:EoMqr}. This equation should be thought of as determining $\hat{q}(0,0)$.

Furthermore, a final constraint is associated with the parameter $m$ counting the number of clones of the system. We are free to analytically continue to any $m\in\mathbb{R}$ and think of it as parameterizing a landscape of states. In an SYK-like `paramagnetic' phase, $m=1$ as replica symmetry is unbroken and clones remain uncorrelated. More generally, however, we wish to choose whichever value dominates dynamically. This turns out to be achieved by the following consideration: the eigenvalues of the Hessian of the overlap matrix $Q_{ab}$ characterize the stability of any given configuration; these depend on $m$. Positive eigenvalues correspond to stable states, while negative ones are unstable. The dynamics prefers states that are {\it marginally stable}, i.e., right at the onset of slow relaxation, where soft modes exist \cite{PhysRevB.36.5388,crisanti1992sphericalp,PhysRevLett.71.173,de1995static}. Therefore, marginal stability corresponds to the value of $m$ where the Hessian has one vanishing eigenvalue. This turns out to correspond to the following condition:
\begin{equation}\label{eq:replicon}
    \frac{1}{\left(\frac{\hq(0,0)}{\beta L}-u\right)^2}=\frac{p(p-1)}{2}(\beta L J)^2u^{p-2}\,.
\end{equation}
Combining it with \eqref{eq:usol}, we can obtain a simple relation for the parameter $m$:
\begin{equation}
\label{eq:marginalm}
    m = \frac{p-2}{\beta L \mathcal{J} \, u^{p/2}}\,,\qquad \mathcal{J}\equiv J\sqrt{\frac{p(p-1)}{2}}\,.
\end{equation}

\paragraph{Thermodynamic potential and free energy.}

The thermodynamic properties of the model are encoded in the "grand-canonical" potential $\overline{\Phi}$ defined in \eqref{eq:PhiDef}. In the large-$N$ limit, it can be obtained from the effective action \eqref{eq:Seff} as 
\begin{equation}\label{eq:PhiderivSeff}
    \beta m \overline{\Phi}=\lim_{n\to 0}\partial_n\Seff(Q_\star)\,,
\end{equation}
where $Q_\star$ denotes a saddle-point solution.

The effective action \eqref{eq:Seff} evaluated on-shell contains ultraviolet divergences, which need to be regularized. We do this in detail in Appendix \ref{app:FreeEnReg}. The final result is
\begin{equation}\label{eq:Seffreg}
\begin{aligned}
    \frac{\Seff^{\rm reg}}{nmN}=&-\frac{1}{2}\sum_{k,l}\log \left\{ \left[\left(\frac{2\pi k}{\beta}\right)^2 + \left(\frac{2\pi l}{L}\right)^2+\lambda\right]\hat{q}_r(k,l)\right\} + \frac{1}{2m}\log \left[\frac{\frac{\hat{q}_r(0,0)}{\beta L}}{\frac{\hat{q}_r(0,0)}{\beta L}+ m u}  \right]\\
    & -\frac{1}{2} \sum_{k,l} \left\{1-\left[\left(\frac{2\pi k}{\beta}\right)^2 + \left(\frac{2\pi l}{L}\right)^2 +\lambda \right]\hat{q}_r(k,l) \right\}+\frac{\beta L\lambda \,u}{2} \\
    & - \frac{(\beta LJ)^2 }{4} \left[ (m-1)u^p + \frac{1}{\beta L} \int_\mathcal{M} d^2\bd{x}\ q(\bd{x})^p \right] \\
    &-\beta L\left[\frac{\lambda}{8\pi}\log\frac{\lambda}{e\Lambda^2}+\frac{1}{2\pi}\sum_{(k,l)\neq (0,0)}\sqrt{\frac{\lambda}{\left((k\beta)^2+(lL)^2\right)}}K_1\left(\sqrt{\lambda ((k\beta)^2+(lL)^2)}\right)\right]  \,.
\end{aligned}
\end{equation}

Since the regularized effective action is proportional to $n$, from \eqref{eq:PhiderivSeff} we readily obtain the grand-canonical potential density $\overline{\phi}\equiv\overline{\Phi}/N$ as
\begin{equation}\label{eq:Phidensity}
    \beta m \overline{\phi}=\frac{1}{N}\lim_{n\to 0}\partial_n\Seff^{\rm reg}(Q_\star)=\frac{\Seff^{\rm reg}}{nN}\,.
\end{equation}
In the ensemble under consideration, where $m$ is treated as an external parameter, the free energy is given by \cite{zamponi2010mean}
\begin{equation}
    \beta\bar{f}=\partial_m(m\beta \overline{\phi})=\beta \overline{\phi}+m\partial_m(\beta\overline{\phi})\,.
    \label{eq:betaf}
\end{equation}
The partial derivative in the second term can be computed explicitly, leading to
\begin{equation}\label{eq:betafdensity}
    \beta \bar{f}=\beta \overline{\phi}-\frac{m}{2} \left[\frac{(\beta LJ)^2}{2}u^p+\frac{1}{m}\frac{u}{\frac{\hat{q}_r(0,0)}{\beta L}+ m u}+\frac{1}{m^2}\log\left[\frac{\frac{\hat{q}_r(0,0)}{\beta L}}{\frac{\hat{q}_r(0,0)}{\beta L}+ m u}\right]\right]\,.
\end{equation}
Once the free energy is obtained using \eqref{eq:betafdensity}, the other thermodynamic quantities follow as
\begin{equation}
    \overline{e}=\partial_\beta(\beta \bar{f})\,,\qquad \overline{s}=\beta(\overline{e}-\bar{f})\,.
\end{equation}

\subsection{Scaling structure of the Schwinger-Dyson equations}\label{subsec:scaling}

Before turning to a detailed analysis of the equations of motion \eqref{eq:EoMqr}-\eqref{eq:marginalm} and their numerical solutions, it is useful to summarize the potential qualitative behavior of the system across parameter space. To this end, we will now explore the possibility of scaling solutions at large values of $\beta$, $L$ and $J$. Such an analysis turns out to be instructive both in the paramagnetic phase and in the spin glass.

\paragraph{Paramagnetic regime.}
When the disorder strength is sufficiently weak, the self-energy in \eqref{eq:EoMqr} provides only a small correction to the propagator. In this regime the dynamics are dominated by the kinetic term and the mass generated by the constraint. The two-point function therefore behaves similarly to that of the $O(N)$ model, cf.\ Fig. \ref{fig:CPN-1}, or the analytical result \eqref{eq:qposplanar} in the planar limit, $\beta,L\to\infty$. In all cases, the inverse zero mode is non-zero, which translates into an exponential decay of the position correlator at large separations.

At strong disorder or large $(\beta, L)$, the self-energy can dominate against the kinetic term. In this situation it is natural to explore the possibility of scaling solutions of the Schwinger–Dyson equations.

To this end, we employ the usual trick of dropping the kinetic term in the Schwinger-Dyson equation \eqref{eq:EoMqr}, i.e,
\begin{equation}
    \frac{1}{\hat{q}_r(k,l)}-\frac{1}{\hat{q}_r(0,0)} \approx -J^2\left(\hat{\Lambda}_r(k,l)-\hat{\Lambda}_r(0,0)\right)\,.
    \label{eq:SDeqIR}
\end{equation}
We can consider the planar limit, $\beta,L\to\infty$ and look for solutions where the two-point function $q(\bd{x})$ behaves as
\begin{equation}
    q^c(\bd{x})=u+q_r^c(\bd{x})\, ,\qquad q_r^c(\bd{x})=\frac{A}{\abs{\bd{x}}^{2\Delta}}\,,
    \label{eq:q_r^c}
\end{equation}
where "c" stands for conformal. In momentum space, this becomes
\begin{equation}
    \hat{q}_r^c(\bd{k})=\int d^2\bd{x}\, q_r^c(\bd{x})e^{i\bd{k}\cdot\bd{x}}=\frac{A\,\pi}{2^{2\Delta-2}}\frac{\Gamma(1-\Delta)}{\Gamma(\Delta)}\abs{\bd{k}}^{2\Delta-2}\,,
    \label{eq:confqmom}
\end{equation}
where in this limit, momentum becomes a continuum variable. The self-energy is given by
\begin{equation}
    \Lambda_r(\bd{x})=\frac{p}{2}q_r^c(\bd{x})^{p-1}=\frac{p}{2}\frac{A^{p-1}}{\abs{x}^{2\Delta(p-1)}}\,
\end{equation}
which, in momentum space, is
\begin{equation}
    \hat{\Lambda}_r^c(\bd{k})=\int d^2\bd{x}\, \Lambda_r^c(\bd{x})e^{i\bd{k}\cdot\bd{x}}=\frac{\pi pA^{p-1}}{2^{2\Delta(p-1)-1}}\frac{\Gamma(1-\Delta(p-1)}{\Gamma(\Delta(p-1))}\abs{\bd{k}}^{2\Delta(p-1)-2} \,.
    \label{eq:confLambdamom}
\end{equation}

The equation of motion \eqref{eq:SDeqIR} contains the zero modes $1/\hat{q}_r(0,0)$ and $\hat{\Lambda}_r(0,0)$. Notice that, from \eqref{eq:confqmom}, $1/\hq_r^c(\bd{0})\to 0$ if $\Delta<1$. Equivalently, from \eqref{eq:confLambdamom}, $\hL_r^c(\bd{0})\to 0$ if $\Delta>\frac{1}{p-1}$. We can drop for now the zero modes and check later that the value we obtain for $\Delta$ is consistent with these assumptions. Eq. \eqref{eq:SDeqIR} then reduces to
\begin{equation}
    \frac{1}{\hat{q}_r^c(\bd{k})}=-J^2\hat{\Lambda}_r^c(\bd{k})\,.
\end{equation}
Using \eqref{eq:confqmom} and \eqref{eq:confLambdamom}, a solution exists provided that
\begin{equation}
    \Delta=\frac{2}{p},\qquad A^pJ^2 = \frac{2}{p}\frac{(1-\Delta)^2}{\pi^2}\,.
    \label{eq:DeltaandA}
\end{equation}
The result $\Delta=2/p$ is consistent with the two inequalities when $p>2$.

We can use a conformal map to obtain the conformal ansatz \eqref{eq:q_r^c} on the cylinder ($L\to\infty$ with finite $\beta$):
\begin{equation}
    q_{r,c}^{cyl}(\tau,x)=A\left[\frac{2\left(\pi/\beta\right)^2}{\cosh\frac{2\pi}{\beta}x-\cos\frac{2\pi}{\beta}\tau}\right]^{2/p}\,.
    \label{eq:conformalParcylinder}
\end{equation}

Such scaling solutions are analogous to the conformal regime of the SYK model or the CFT${}_2$ conjectured in the models of \cite{Murugan:2017eto}.

\paragraph{Spin glass regime.}
As mentioned above, disordered bosonic models tend to undergo a transition to a spin glass at low temperatures and strong coupling. In this case, replica symmetry is explicitly broken and the Edwards–Anderson parameter $u$ becomes non-zero. These are precisely the limits where the potential scaling solution would emerge, and therefore a full numerical solution of the model is required to determine which type of solution is in practice realized.

This regime corresponds to solutions with
\begin{equation}
    q(\bd{x}) = u + q_r(\bd{x})\,, \qquad u \neq 0\,.
\end{equation}
It is natural to ask whether a scaling ansatz for the non-constant part $q_r(\xv)$ can be realized within the spin glass phase.

In order to do so, it is instructive to consider the solutions in the "deep spin glass" regime, corresponding to $q_r(\xv)\ll u$.
We begin by considering the Schwinger-Dyson equation \eqref{eq:EoMqr} in position space (equivalently, the diagonal entries of \eqref{eq:Schwinger-Dyson}):
\begin{equation}
    -\delta^{(2)}(\xv,\xvp)+\frac{q_r(\xv,\xvp)}{\hq_r(0,0)} =\partial_\mu^2 q_r(\xv,\xvp)+{J^2} \intxpp\Lambda_r(\bd{x},\bd{x}'')\left[q_r(\bd{x}'',\bd{x}')-q_r(\bd{x},\bd{x}')\right]\,,
\end{equation}
with
\begin{equation}
    \Lambda_r(\bd{x}, \bd{x}') \equiv \frac{p}{2}\left[\left(q_r(\bd{x},\bd{x}')+u\right)^{p-1} - u^{p-1}\right]\,.
\end{equation}
When $q_r(\xv)\ll u$, we can expand $\Lambda_r(\xv, \xvp')$ in powers of $q_r(\xv,\xvp)$:
\begin{multline}\label{eq:eompowers}
    -\delta^{(2)}(\xv,\xvp)+\frac{q_r(\xv,\xvp)}{\hq_r(0,0)} =\partial_\mu^2 q_r(\xv,\xvp)\\+{\mathcal{J}^2}u^{p-2} \intxpp\left[q_r(\xv,\xvpp)+\frac{p-2}{2u}q_r(\xv,\xvpp)^2+\dots\right]\!\left[q_r(\bd{x}'',\bd{x}')-q_r(\bd{x},\bd{x}')\right]\,,
\end{multline}
where $\mathcal{J}$ was defined in \eqref{eq:marginalm}.

By applying the conditions for marginal stability, \eqref{eq:replicon}-\eqref{eq:marginalm}, we can write the zero mode in terms of the coupling $\mathcal{J}$,
\begin{equation}
\label{eq:zeromodeJu}
    \frac{1}{\hq_r(0,0)^2}=\mathcal{J}^2u^{p-2}\,.
\end{equation}
Using this relation and keeping only the leading term in $q_r(\xv,\xvp)$ the equation of motion \eqref{eq:eompowers} can be written as
\begin{equation}
    -\partial_\mu^2 q_r(\xv,\xvp) =\intxpp\left[\delta^{(2)}(\xv,\xvpp)-\frac{q_r(\xv,\xvpp)}{\hq_r(0,0)}\right]\left[\delta^{(2)}(\xvpp,\xvp)-\frac{q_r(\xvpp,\xvp)}{\hq_r(0,0)}\right]\,,~\label{eq:eommassaged}
\end{equation}
which, if we assume translation invariance, takes a simple form in momentum space:
\begin{equation}
    \left(\omega_k^2+p_l^2\right)\hq_r(k,l)=\left(\frac{\hq_r(k,l)-\hq_r(0,0)}{\hq_r(0,0)}\right)^2\,,
\end{equation}
with solution
\begin{equation}\label{eq:2Dapproxq}
    \frac{\hq_r^{\approx}(k,l)}{\hq_r(0,0)}=1+2\gamma^2\left(\omega_k^2+p_l^2\right)-2\sqrt{\gamma^2\left(\omega_k^2+p_l^2\right)+\gamma^4\left(\omega_k^2+p_l^2\right)^2}\,,\qquad \gamma\equiv\sqrt{\frac{\hq_r(0,0)}{4}}\,.
\end{equation}
In the limit $(\beta,L)\to\infty$, where frequencies and momenta become continuous, the Fourier transform to position space can be obtained exactly:
\begin{equation}
\label{eq:I0K0-I2K2}
    q_r^\approx(\bd{x})=\frac{1}{2\pi} \left[I_0\left(\frac{|\bd{x}|}{2\gamma}\right)K_0\left(\frac{|\bd{x}|}{2\gamma}\right)-I_2\left(\frac{|\bd{x}|}{2\gamma}\right)K_2\left(\frac{|\bd{x}|}{2\gamma}\right)\right]\,.
\end{equation}
It turns out that this expression provides an excellent approximation to the full solution, similarly to the one-dimensional case \cite{Anous:2021eqj}.

We can consider again the momentum-space solution \eqref{eq:2Dapproxq} and expand it for low momenta in the continuum limit,

\begin{equation}
    \frac{\hat{q}^{\approx}_r (\bd{k})}{\hat{q}_r(\bd{0})}=1-2\gamma|\bd{k}|+\ldots\,,
\end{equation}
with Fourier transform
\begin{equation}
    q^\approx_r(\bd{x}) = \hat{q}_r(\bd{0})\delta^{(2)}(\bd{x}) + \frac{A}{|\bd{x}|^3}\,,\qquad A=\frac{4\gamma^3}{\pi}\,,
    \label{eq:SGconformal}
\end{equation}
consistent with the scaling of a conformal operator of effective dimension $\Delta=3/2$.

We can apply a conformal mapping to obtain the corresponding conformal ansatz on the cylinder:
\begin{equation}
    q^c_r(\tau,x) =  A \left[\frac{2 (\pi/\beta)^2}{\cosh \left(\frac{2\pi x}{\beta}\right)-\cos\left(\frac{2\pi\tau}{\beta}\right)}\right]^{3/2}\,.
    \label{eq:conformalSGcylinder}
\end{equation}

Determining which of these possibilities is realized requires a numerical solution of the full Schwinger–Dyson equations, including the contribution of the zero modes, which is the task of the following sections.

\section{The static approximation}\label{sec:staticanalysis}

As a first step towards understanding the phase structure of the disordered $O(N)$ non-linear sigma model, we generalize the so-called "static" approximation \cite{Bray_moore,crisanti1992sphericalp,Cugliandolo_2001} to our field theory. This approximation consists of assuming that the bilocal fields are constant,
\begin{equation}
    Q_{ab}(\bd{x}) \to Q_{ab} = \text{constant}\,,\quad \Lambda_{ab}(\bd{x})\to\Lambda_{ab} = \text{constant}\, ,
\end{equation}
which in momentum space implies
\begin{equation}
    \hat{Q}_{ab}(k,l) =\beta L\, Q_{ab}\,\delta_{k,0}\delta_{l,0}\,,
\end{equation}
and similarly for $\Lambda_{ab}$. This reduces the full Schwinger-Dyson equations to a finite dimensional saddle-point problem. Although this approximation neglects the spacetime structure of the correlators, it often provides a sharp diagnosis of the organization of solutions and the phase diagram. We verify this a posteriori.

Throughout the numerics in this section we restrict attention to the case $p=3$, which already captures all qualitative features of the model. Nevertheless, we keep $p$ explicit in the equations whenever possible.

We first need to evaluate the effective action \eqref{eq:SefflargeN} in this approximation, which yields
\begin{equation}
\begin{aligned}
    \frac{S_\text{eff}}{N} =\,& -\frac{1}{2}\sum_{a,b}\log \left(\beta L\,Q_{ab}\right)+\frac{nm}{2}\sum_{(k,l)\neq(0,0)}\log\!\left[\omega_k^2+p_l^2+\lambda\right]-\frac{(\beta LJ)^2}{4}\sum_{a,b}(Q_{ab})^p\\
    & -\frac{nm}{2}\beta L\lambda\left(\frac{1}{g_0^2}-q_d\right)\, ,
  \label{eq:Seffstatic}
\end{aligned}
\end{equation}
where $q_d\equiv Q_{aa}$.

The regularization of the momentum sum and the subtraction of the bare coupling $1/g_0^2$ proceed as in the previous section. The renormalized effective action then becomes
\begin{equation}
\begin{aligned}
    \frac{S_\text{eff}^{\rm reg}}{N} =\,& -\frac{1}{2}\sum_{a,b}\log \left(Q_{ab}\right)-\frac{nm}{2}\log(\beta L\lambda)-\frac{(\beta LJ)^2}{4}\sum_{a,b}(Q_{ab})^p+\frac{nm}{2}\beta L\lambda\,q_d\\
    & -nm\beta L\left[\frac{\lambda}{8\pi}\log\frac{\lambda}{e\Lambda^2}+\frac{1}{2\pi}\sum_{(k,l)\neq (0,0)}\sqrt{\frac{\lambda}{\left((k\beta)^2+(lL)^2\right)}}\,K_1\left(\sqrt{\lambda ((k\beta)^2+(lL)^2)}\right)\right] .
\end{aligned}
\end{equation}
Varying with respect to the Lagrange multiplier $\lambda$ gives the constraint equation
\begin{equation}
    q_d = \frac{1}{\beta L\lambda}+ \frac{1}{4\pi}\log\frac{\lambda}{\Lambda^2}- \frac{1}{2\pi}\sum_{\substack{(k,l)\neq\\(0,0)}}
    K_0\!\left(\sqrt{\lambda\bigl((k\beta)^2+(lL)^2\bigr)}\right)\,,\label{eq:EoMwrtlambda}
\end{equation}
while the variation with respect to $Q_{ab}$ yields
\begin{equation}
    \lambda\,\delta_{ab} = \frac{Q_{ab}^{-1}}{\beta L} + \frac{\beta L}{2}\,p\,J^2\,Q_{ab}^{p-1} \,.
    \label{eq:EoMwrtQab}
\end{equation}

Equations \eqref{eq:EoMwrtlambda} and \eqref{eq:EoMwrtQab} are the central saddle-point equations of the static analysis. They are structurally similar to those of the quantum spherical $p$-spin model studied in \cite{Cugliandolo_2001}. In particular, the replica equation \eqref{eq:EoMwrtQab} is identical to its one-dimensional counterpart, while the constraint equation \eqref{eq:EoMwrtlambda} differs due to the distinct UV structure of the two-dimensional theory.

The grand-canonical potential density $\overline{\phi}$ is given by \eqref{eq:Phidensity},
\begin{equation}
\begin{aligned}
    \beta \overline{\phi} &=-\lim_{n\to 0}\partial_n\sum_{a,b}\frac{1}{2m}\left[\log \left(Q_{ab}\right)+ \frac{(\beta LJ)^2}{2}(Q_{ab})^p \right]-\frac{1}{2}\log(\beta L\,\lambda)+\frac{1}{2}\beta L\lambda\,q_d\\
    &\quad -\beta L \left[\frac{\lambda}{8\pi}\log\frac{\lambda}{e\Lambda^2}+\frac{1}{2\pi}\sum_{(k,l)\neq (0,0)}\sqrt{\frac{\lambda}{\left((k\beta)^2+(lL)^2\right)}}K_1\left(\sqrt{\lambda ((k\beta)^2+(lL)^2)}\right)\right]\,.
    \label{eq:phistatic}
\end{aligned}
\end{equation}

A comment is in order: in the paramagnetic phase, free energy considerations provide the appropriate criterion for distinguishing stable and unstable branches, as well as which stable branch dominates. In the spin glass phase, however, the parameter $m$ is determined from the condition of marginal stability \eqref{eq:replicon}-\eqref{eq:marginalm}, which is not necessarily a minimum of the free energy. Consequently, free energies will be used as a diagnostic tool for characterizing the different branches, while the onset of glassy behavior will be determined by the existence of marginal solutions rather than by a free energy minimization.

We now analyze the different classes of saddle-point solutions to Eqs. \eqref{eq:EoMwrtlambda}, \eqref{eq:EoMwrtQab}.

\subsection{Paramagnetic solution}
The simplest saddle corresponds to the paramagnetic ansatz,
\begin{equation}
    Q_{ab}(\bd{x})=q_d\,\delta_{ab}\,.
\end{equation}
Equation \eqref{eq:EoMwrtQab} then reduces to
\begin{equation}
    \lambda=\frac{1}{\beta L\,q_d}+\frac{\beta L}{2}pJ^2q_d^{p-1}\,.
    \label{eq:lambdaPM}
\end{equation}
Substituting \eqref{eq:lambdaPM} into the constraint equation \eqref{eq:EoMwrtlambda} determines the allowed paramagnetic saddles. To study these solutions it is convenient to define the function
\begin{equation}
    g(q_d)=q_d-\text{rhs}\left[\text{Eq.}\,\eqref{eq:EoMwrtlambda}\right]\,,
    \label{eq:g(qd)}
\end{equation}
whose zeros determine the different solutions.

\begin{figure}
    \centering
    \includegraphics[width=0.45\textwidth]{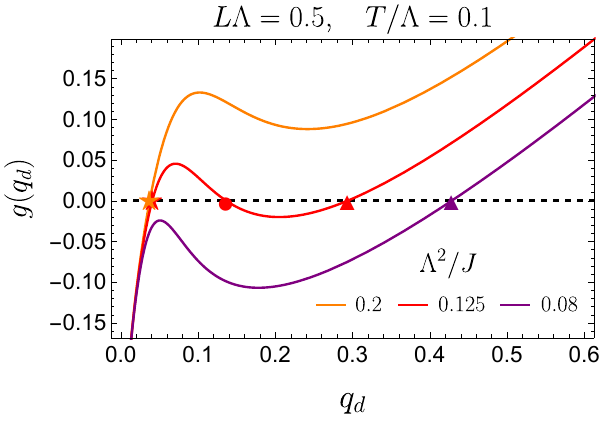}
    \includegraphics[width=0.45\textwidth]{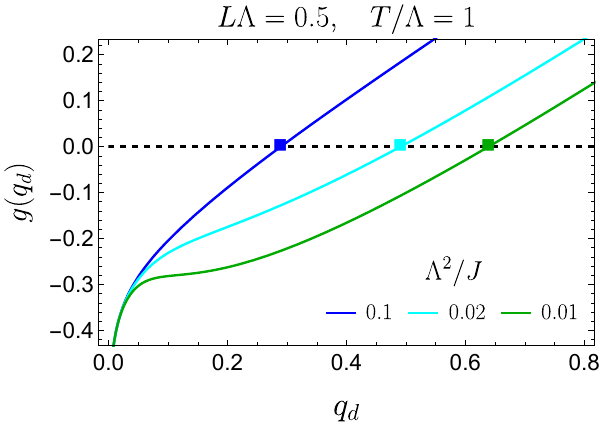}\\
    \includegraphics[width=0.45\textwidth]{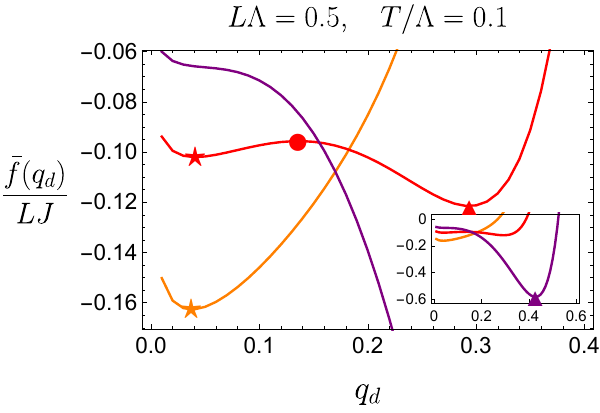}
    \includegraphics[width=0.45\textwidth]{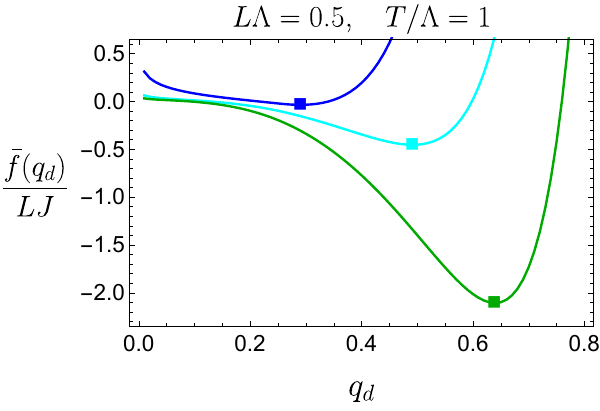}
    \caption{\textit{Top:} Paramagnetic solutions are obtained as the zeros of the function $g(q_d)$, defined in \eqref{eq:g(qd)}. The left (right) panels correspond to low (high) temperatures for fixed $L\Lambda=0.5$. At high temperatures, the solution is unique (indicated by squares). Below a critical temperature $T_{\rm PM}$ there exists a range of couplings where three paramagnetic solutions coexist (left panel, red curve). In this case, stars correspond to the PM$_g$ solutions, triangles are the PM$_{th}$ solutions, and the circle denotes the PM$_u$ solution. \textit{Bottom:} Free energies associated with the corresponding solutions. The inset in the left plot shows a zoom out away from the coexistence region. When three solutions coexist, only two correspond to local minima of the free energy, while the intermediate solution PM$_u$ corresponds to a local maximum of the free energy and is therefore unstable.}
    \label{fig:qdsolsPM}
\end{figure}

The resulting structure is shown in Fig. \ref{fig:qdsolsPM} (upper plots). At high temperatures there exists a unique paramagnetic saddle for all values of the coupling. However, below a critical temperature $T_{\rm PM}$, a coexistence region emerges in which three distinct paramagnetic solutions coexist.

The corresponding free energies (lower plots) reveal that only two of these branches are locally stable. We denote them by\footnote{The names {\it gapped} and {\it thermal} are related to their behavior in the full numerical solution. We will see in Figs. \ref{fig:PM1st2pf} and \ref{fig:PM1stfzm} that PM$_g$ has a nontrivial mass scale while PM$_{th}$ has thermal fluctuations that dominate over quantum fluctuations. In the one-dimensional static analysis of \cite{Cugliandolo_2001}, the analogue of PM$_g$ was referred to as the quantum paramagnet (PM$_1$) while the analogue of PM$_{th}$ was referred to as the classical paramagnet (PM$_2$).}
\begin{itemize}
    \item PM$_g$, the \emph{gapped paramagnet}, which exists at relatively weak coupling, and 
    \item PM$_{th}$, the \emph{thermal paramagnet}, which dominates at stronger coupling.
\end{itemize}
The remaining branch, denoted PM$_u$, corresponds to a local maximum of the free energy and is therefore unstable.

The coexistence of PM$_g$ and PM$_{th}$, separated by the unstable branch PM$_u$, signals a first-order transition between the two locally stable paramagnetic phases. Within the coexistence region, the free energies of PM$_g$ and PM$_{th}$ cross, leading to an exchange of thermodynamic dominance (see Fig. \ref{fig:freeEnPMSG} below).

Repeating the analysis for different values of $L\Lambda$ changes the quantitative details but leaves the qualitative picture unchanged. In particular, increasing $L\Lambda$ shifts the solutions towards smaller values of $q_d$.

An important feature inherited from the one-dimensional model is that the PM$_{th}$ branch, although stable from the viewpoint of the static free energy, is associated with a pathological solution at low temperatures \cite{Cugliandolo_2001}. The thermal paramagnet exhibits divergent thermodynamic quantities as $T\to 0$, and is therefore regarded as unphysical \cite{Cugliandolo_2001, Anous:2021eqj}. The same qualitative behavior appears to hold in the present two-dimensional model.

In order to allow for the possibility of a spin glass phase at strong coupling, we now consider the static approximation for the 1-RSB ansatz.

\subsection{1-step replica symmetry-breaking (1-RSB) solution}
The static approximation for the 1-RSB ansatz given in \eqref{eq:1rsb} amounts to replacing the diagonal entry $q(\xv,\xvp)$ by $q_d$. Substituting the static 1-RSB ansatz into the saddle-point equation \eqref{eq:EoMwrtQab}, inverting the matrix, and taking the replica limit $n\to0$, one obtains
\begin{align}
    \beta L\lambda&=\frac{q_d+(m-2)u}{\left[q_d+u(m-1)\right](q_d-u)}+\frac{(\beta L J)^2}{2}\,p \, q_d^{p-1}\,,\label{eq:1RSBeqlambda}\\
    0&=\frac{1}{\left[q_d+u(m-1)\right](q_d-u)}-\frac{(\beta L J)^2}{2}\,p\,u^{p-2}\,.\label{eq:1RSBequ}
\end{align}
These two equations are supplemented by the constraint equation \eqref{eq:EoMwrtlambda} and the marginality condition \eqref{eq:marginalm}, which together completely characterize the possible marginal spin glass solutions.\footnote{At this point it is important to emphasize the conceptual difference with the equilibrium 1-RSB construction. In the equilibrium solutions (not discussed here, see \cite{Cugliandolo_2001,Anous:2021eqj} for a detailed analysis in the one-dimensional model) the breakpoint parameter $m$ is determined by extremizing the free energy. Here we instead focus on marginal states and impose the condition \eqref{eq:marginalm}. The resulting spin-glass solutions should therefore be interpreted as marginally stable dynamical states rather than equilibrium configurations. Accordingly, the physically relevant phase boundaries will be determined by the existence of solutions satisfying the marginality condition rather than by comparisons of free energies between competing saddles.}

The free energy associated with these solutions is nevertheless useful because it provides additional information about the structure of the different branches and facilitates comparison with previous studies of related models. The grand-canonical potential is now directly obtained from \eqref{eq:Seffreg}-\eqref{eq:Phidensity} and is given by
\begin{equation}
\begin{aligned}
    \beta \overline{\phi} &= -\frac{1}{2}\left(1-\frac{1}{m}\right)\log(q_d-u) -\frac{1}{2m}\log\bigl(q_d+(m-1)u\bigr)\\
    &\quad -\frac{(\beta LJ)^2}{4}\left(q_d^p+(m-1)u^p\right)-\frac{1}{2}\log(\beta L\lambda)+\frac{1}{2}\beta L\lambda\,q_d\\
    &\quad -\beta L\left[\frac{\lambda}{8\pi}\log\frac{\lambda}{e\Lambda^2}+\frac{1}{2\pi}\sum_{(k,l)\neq (0,0)}\sqrt{\frac{\lambda}{\left((k\beta)^2+(lL)^2\right)}}K_1\left(\sqrt{\lambda ((k\beta)^2+(lL)^2)}\right)\right]\,,
      \label{eq:betaPhi1RSB}
\end{aligned}
\end{equation}
and the free energy is again
\begin{equation}\label{eq:betaf1RSB}
    \beta\bar{f}=\beta\overline{\phi}-\frac{m}{2}\left[\frac{(\beta LJ)^2}{2}u^p+\frac{1}{m}\frac{u}{q_d+u(m-1)}+\frac{1}{m^2}\log\frac{q_d-u}{q_d+u(m-1)}\right]\,.
\end{equation}

We can now solve the marginal equations numerically and determine for which values of the coupling they admit spin-glass solutions. As in the paramagnetic case, the resulting phase structure depends qualitatively on whether the temperature lies above or below a critical temperature $T_c$.

For $T>T_c$, the PM-to-SG transition is continuous. The paramagnetic solution exists for all values of the coupling. At a critical coupling, $J_c$, a marginal spin-glass solution first appears and continuously emerges from the paramagnetic branch. The appearance of this solution signals the onset of glassy order. Solutions with $m>1$ formally exist beyond the transition point but must be discarded since the replica construction requires $0<m\le1$.

For $T<T_c$, the coexistence of PM$_g$ and PM$_{th}$ leads to a richer structure. In this regime marginal spin-glass solutions emerge in the vicinity of the PM$_g$ branch and can be continuously connected to it. By contrast, PM$_{th}$ remains disconnected from the marginal spin-glass solution and corresponds to the continuation of the pathological thermal branch discussed above. Consequently, the physically relevant transition involves PM$_g$ and the onset of marginal glassy order. The resulting spin-glass branch provides the physically relevant continuation beyond the instability of the paramagnetic phase at $J=J_c$.

Repeating the analysis for different values of $L\Lambda$ modifies the location of the transition lines quantitatively but leaves the qualitative structure unchanged.

\begin{figure}
    \centering
    \includegraphics[width=0.45\textwidth]{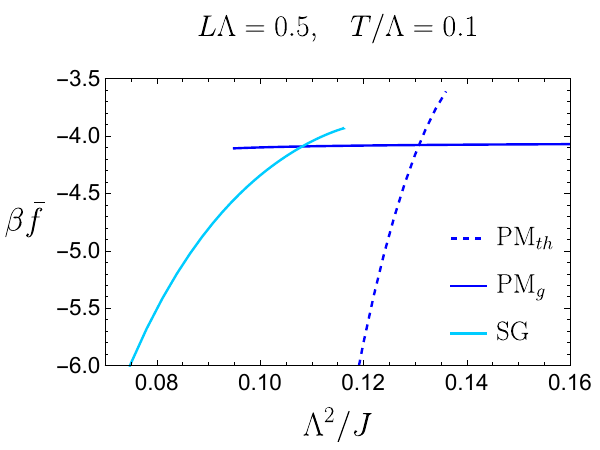}
    \includegraphics[width=0.45\textwidth]{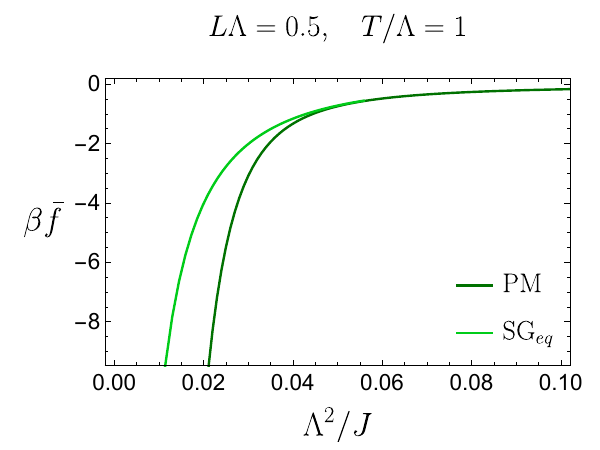}
    \caption{Free energies of the paramagnetic and spin glass solutions at low temperatures (left) and high temperatures (right). At low temperatures, where the PM$_g$ and PM$_{th}$ solutions coexist, an exchange of thermodynamic dominance between these two solutions occurs. However, the PM$_{th}$ are unphysical and the system remains in the PM$_g$ phase until the SG solution begins to exist.}
    \label{fig:freeEnPMSG}
\end{figure}

\begin{figure}
    \centering
    \includegraphics[width=0.55\textwidth]{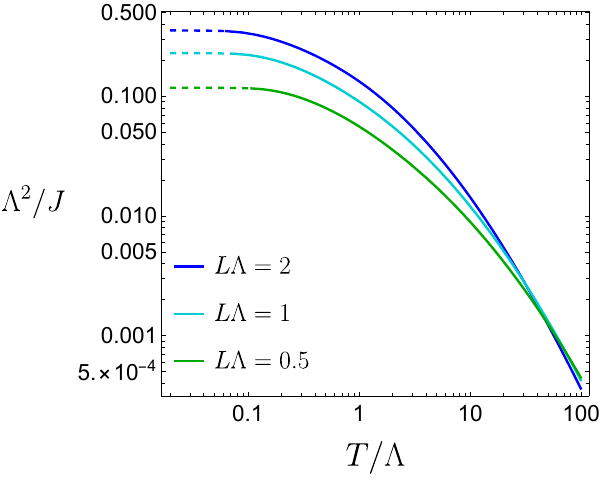}
    \caption{Static phase diagram in the $(J,T)$ plane for different values of $L$. Points above (below) the lines correspond to paramagnetic (spin glass) solutions. Dashed lines correspond to temperatures $T\le T_c$ where the transition is first order. Solid lines are used for $T>T_c$, where the transition is second-order.}
    \label{fig:staticphasediag}
\end{figure}

Putting all these results together, we obtain the phase diagram in the static approximation shown in Fig. \ref{fig:staticphasediag}. We express everything in units of the dynamically generated scale, $\Lambda$. Therefore, the structure of the three-dimensional phase diagram is described in terms of the dimensionless quantities $\beta\Lambda$, $L\Lambda$, which characterize the torus geometry, and $\Lambda^2/J$, measuring the strength of the random interactions. For temperatures below $T_c$, the transition between the paramagnetic and spin-glass phase is first-order and occurs in the presence of the two coexisting phases. Above $T>T_c$, the spin glass solution emerges continuously from the paramagnetic phase, leading to a second-order transition.

Despite its simplicity, the static approximation predicts a remarkably rich phase structure. In particular, it anticipates the coexistence of multiple paramagnetic branches, the emergence of spin-glass order at strong coupling, and a tricritical structure separating first- and second-order transitions. Since the approximation neglects the full spacetime dependence of the correlators, it remains unclear to what extent these features survive in the complete Schwinger-Dyson equations. We address this question in the next section through a direct numerical solution of the full theory.

\section{The numerical solution} \label{sec:numerical}

The static approximation provides a useful qualitative guide, but it neglects the full spacetime structure of the Schwinger–Dyson equations. The present numerical analysis serves two complementary purposes. First, it allows us to determine whether the phase structure predicted by the static approximation survives in the complete theory. Second, it allows us to test whether the scaling solutions identified in Section \ref{subsec:scaling} are dynamically realized in the infrared.

The analytical discussion of Section \ref{sec:disordered} identified a candidate scaling solution with properties reminiscent of the conformal regime encountered in the MSW model \cite{Murugan:2017eto}. However, whether this solution is actually realized by the full Schwinger-Dyson equations is a dynamical question that cannot be answered within the scaling analysis alone or the static approximation of the previous section.

The numerical results presented in this section provide a clear answer. While the theory develops a robust spin-glass phase at sufficiently low temperatures and strong disorder, the replica-symmetric conformal solution is never realized as an infrared fixed point of the full equations. Instead, the numerics reveal a different scenario: the paramagnetic solutions remain gapped throughout the phase diagram, whereas the spin glass solutions show a regime where their behavior agrees with the analytical predictions of the "conformal" deep glassy phase.

We begin by examining the paramagnetic solutions and then turn to the spin-glass phase, where we show the emergence of nontrivial infrared dynamics.

\begin{figure}
    \centering
    \includegraphics[width=\textwidth]{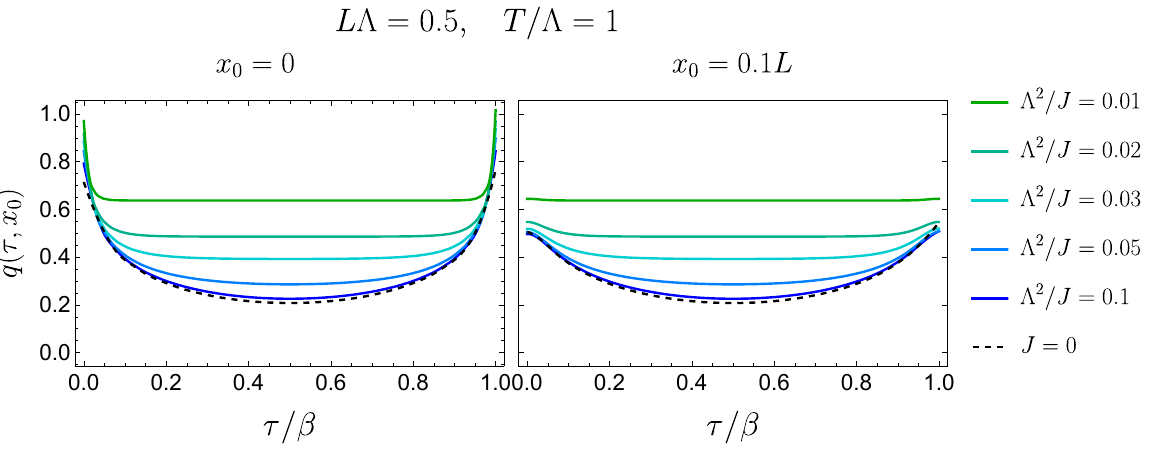}
    \caption{High-temperature regime ($T/\Lambda=1$, $L\Lambda=0.5$). Position-space correlator $q(\tau,x_0)$ at two representative spatial positions $x_0=0$ and $x_0=0.1\,L$, for increasing disorder strength, from bottom to top. The solutions change smoothly with the coupling and are continuously connected to the "free" ($J=0$) solution (dashed).}
    \label{fig:PM2nd2pf}
\end{figure}
\begin{figure}
    \centering
    \includegraphics[width=0.49\textwidth]{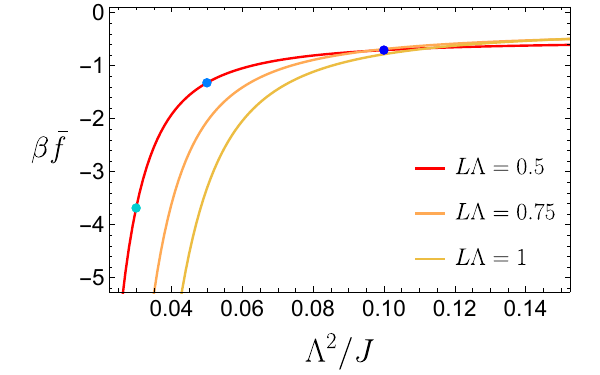}
    \includegraphics[width=0.49\textwidth]{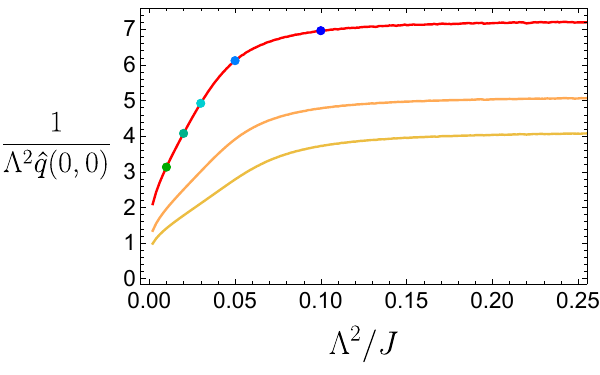}
    \caption{{\it Left}: Free energy as a function of the coupling $\Lambda^2/J$ for the high-temperature paramagnetic solutions, and three values of $L\Lambda$. The dots mark the solutions plotted in Fig. \ref{fig:PM2nd2pf}. {\it Right}: Corresponding inverse zero mode (gap). The gap remains strictly positive for all couplings, confirming the absence of a conformal regime.}
    \label{fig:PM2ndfzm}
\end{figure}

\subsection{The paramagnetic phase: absence of an infrared scaling regime}

We first consider the replica-symmetric solutions, $(u,m)=(0,1)$, which describe the paramagnetic phase.

Motivated by the structure found in the static approximation, it is convenient to fix the size of the torus and vary the disorder strength $J$. As in the static approximation, two qualitatively different temperature regimes must be distinguished.

\paragraph{High temperatures: $\bd{T>T_c}$.} For temperatures above the critical temperature, the Schwinger-Dyson equations admit a unique paramagnetic solution for every value of the coupling. A representative set of position-space correlators $q(\xv)$ is shown in Fig. \ref{fig:PM2nd2pf}.

The corresponding gap and free energy are displayed in Fig. \ref{fig:PM2ndfzm}. As the disorder strength is increased, the gap decreases continuously but remains positive, which implies that no scale-invariant behavior is observed. The infrared dynamics therefore remains controlled by a finite gap throughout the high-temperature paramagnetic phase.

\begin{figure}
    \centering
    \includegraphics[width=\textwidth]{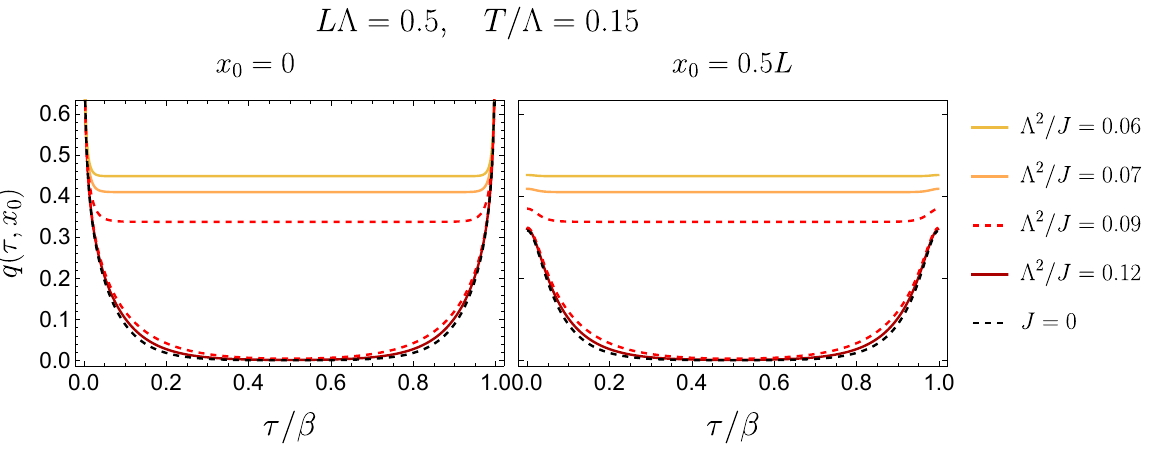}
    \caption{Low-temperature regime ($T/\Lambda=0.15$, $L\Lambda=0.5$). Position-space correlator $q(\tau,x_0)$ at two representative spatial positions $x_0=0$ and $x_0=0.5\,L$, for increasing disorder strength, from bottom to top. The lower curves correspond to the PM$_g$ solutions, while the upper curves correspond to PM$_{th}$. The {\it gapped} paramagnet solutions have a larger gap than the {\it thermal} ones (see Fig. \ref{fig:PM1stfzm}), where the zero mode dominates over the non-zero modes, indicating that thermal fluctuations dominate over quantum fluctuations. The coexistence region is depicted with the two solutions in dashed, sharing the same value of the coupling but belonging to different phases.}
    \label{fig:PM1st2pf}
\end{figure}

\begin{figure}
    \centering
    \includegraphics[width=0.49\textwidth]{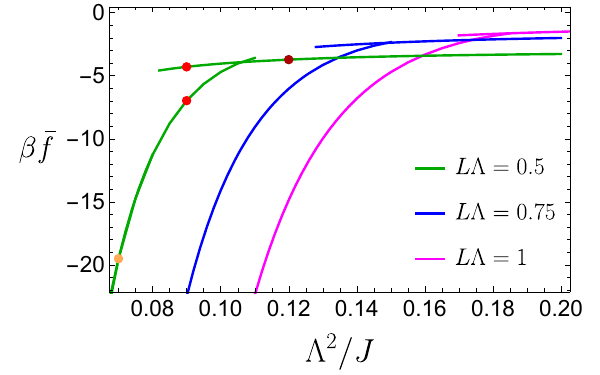}
    \includegraphics[width=0.49\textwidth]{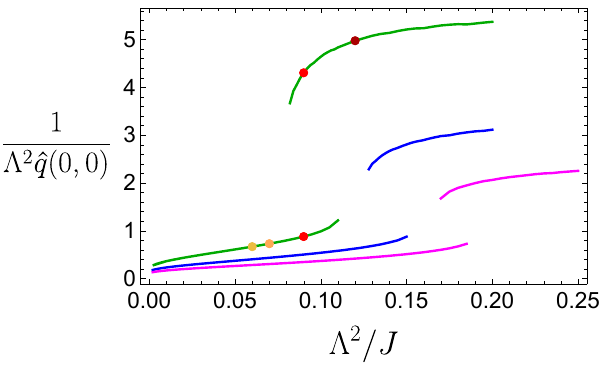}
    \caption{{\it Left}: Free energy as a function of the coupling $\Lambda^2/J$ for the low-temperature paramagnetic solutions, and three values of $L\Lambda$. In the coexistence region, the free energies cross, signalling a first-order transition. However, since PM$_{th}$ is unphysical (see main text), the system remains in PM$_g$ until it ceases to exist (or the spin glass transition takes place). Again, the dots mark the solutions plotted in Fig.\ \ref{fig:PM1st2pf}. {\it Right}: Corresponding inverse zero mode (gap). The PM$_g$ branch (upper curve) is always gapped. The PM$_{th}$ branch (lower curve) has a smaller gap that decreases with increasing $J$.}
    \label{fig:PM1stfzm}
\end{figure}

\paragraph{Low temperatures: $\bd{T<T_c}$.} As in the static approximation, the low-temperature regime is more intricate. As shown in Figs. \ref{fig:PM1st2pf} and \ref{fig:PM1stfzm}, two distinct paramagnetic branches coexist over a finite interval of couplings. The first branch is continuously connected to the "free" ($J=0$) solution and is the dynamical analog of the solution we referred to as PM$_g$ in the static approximation. It exists only up to a maximal coupling $J_g$ (see Fig. \ref{fig:PM1stfzm}). The second branch, denoted by PM$_{th}$, appears only above a threshold coupling $J_{th}$ and extends to arbitrarily large disorder $J$.

The evolution of the corresponding gaps and free energies is summarized in Fig. \ref{fig:PM1stfzm}. In the PM$_g$ branch the gap remains finite and the correlators show a very pronounced decay. In contrast, the PM$_{th}$ branch is characterized by a much smaller gap, which decreases monotonically as the disorder strength is increased and appears to vanish in the limit $J\to\infty$.

A direct inspection of the corresponding two-point functions reveals, however, that the approach to a vanishing gap is not accompanied by the emergence of the scale-invariant behavior discussed in Section \ref{subsec:scaling}. Instead of developing a power-law decay, the correlators become increasingly dominated by a $\tau$-independent plateau. Equivalently, in momentum space the growth of the zero mode $\hq_r(0,0)$ is not accompanied by the power-law dependence on non-zero momenta predicted by the scaling solution. The infrared solution therefore accumulates almost entirely in the zero mode, while the non-zero momentum modes remain subleading.

This observation illustrates that a small, or even vanishing, gap is only a necessary condition for the realization of the conformal solution. The infrared ansatz of Section \ref{subsec:scaling} requires not only a divergent zero mode but also a specific scale-invariant momentum dependence of the non-zero modes. The numerical solutions fail to exhibit this second feature. The vanishing gap is thus associated with the dominance of the zero mode rather than with the onset of a conformal regime.

From this perspective, the PM$_{th}$ solutions should be interpreted as approaching a zero-mode-dominated infrared phase rather than the conformal saddle of the infrared Schwinger-Dyson equations. Consequently, the replica-symmetric sector provides no evidence that the putative conformal regime of the model is dynamically realized.

The coexistence of PM$_g$ and PM$_{th}$ is reminiscent of the structure encountered in the static approximation and the one-dimensional $p$-spin model \cite{Cugliandolo_2001, Anous:2021eqj}, where a phase transition PM$_g\to$\,PM$_{th}$ would occur at the coupling where the free energies cross. However, for the reasons mentioned in the previous section, we interpret the PM$_{th}$ solutions as unphysical and only the PM$_g$ solutions should be considered.

\subsection{Spin glass phase and emergent scaling}

We now allow for replica-symmetry breaking solutions and solve the full set of equations \eqref{eq:EoMqr}-\eqref{eq:marginalm} with non-vanishing Edwards-Anderson parameter, $u\neq 0$.

\begin{figure}
    \centering
    \includegraphics[width=\textwidth]{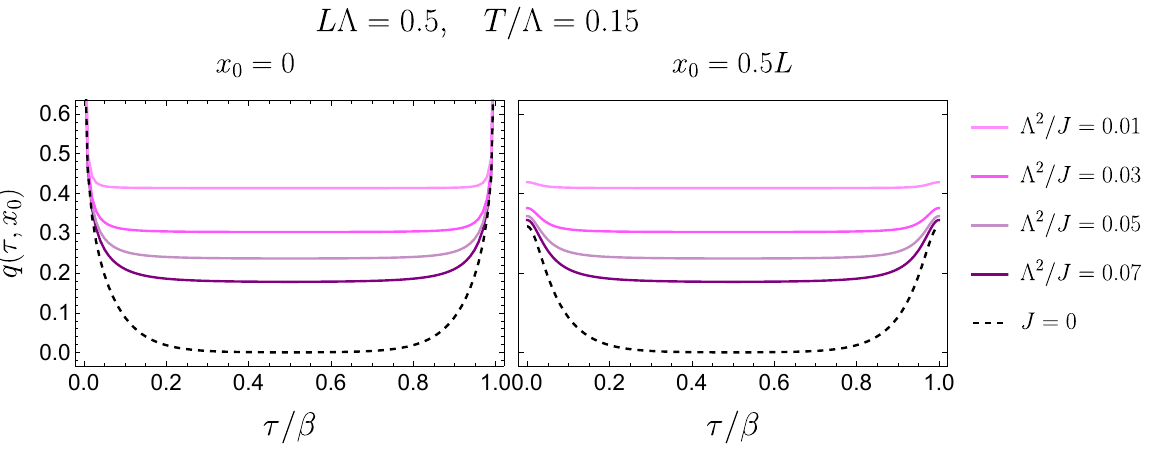}
    \caption{Position-space correlator $q(\tau,x_0)$ at two representative spatial positions $x_0=0$ and $x_0=0.5\,L$, for increasing disorder strength, from bottom to top.}
    \label{fig:SG2pf}
\end{figure}

Spin glass solutions appear below a critical temperature and above a sufficiently strong disorder strength. We show representative correlators in Fig. \ref{fig:SG2pf}.

\begin{figure}
    \centering
    \includegraphics[width=0.49\textwidth]{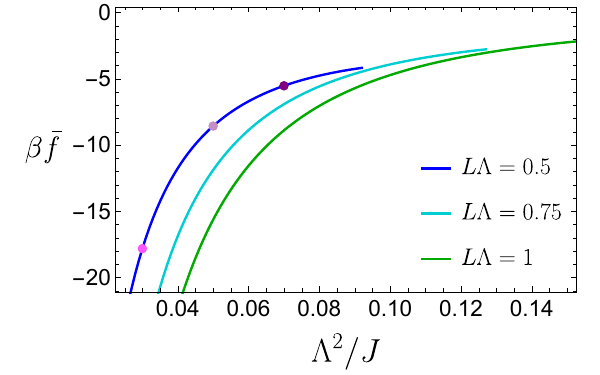}
    \includegraphics[width=0.49\textwidth]{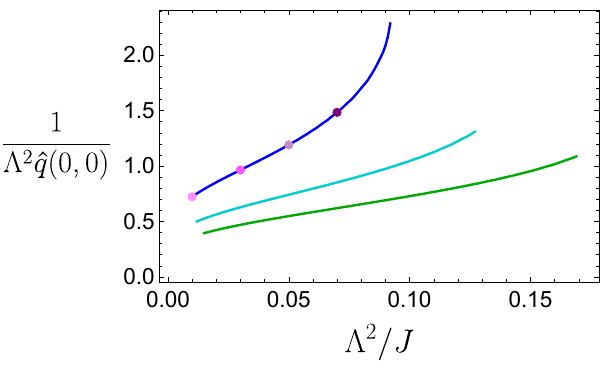}
    \caption{{\it Left}: Free energy as a function of the coupling $\Lambda^2/J$ for the spin glass solutions, and three values of $L\Lambda$. Again, the dots mark the solutions plotted in Fig. \ref{fig:SG2pf}. {\it Right}: Corresponding inverse zero mode (gap).}
    \label{fig:zmfSG}
\end{figure}

The associated gap and free energy are shown in Fig. \ref{fig:zmfSG}. Upon entering the spin glass phase, the gap decreases and the spin glass becomes the only physical solution.

\begin{figure}
    \centering
    \includegraphics[width=\textwidth]{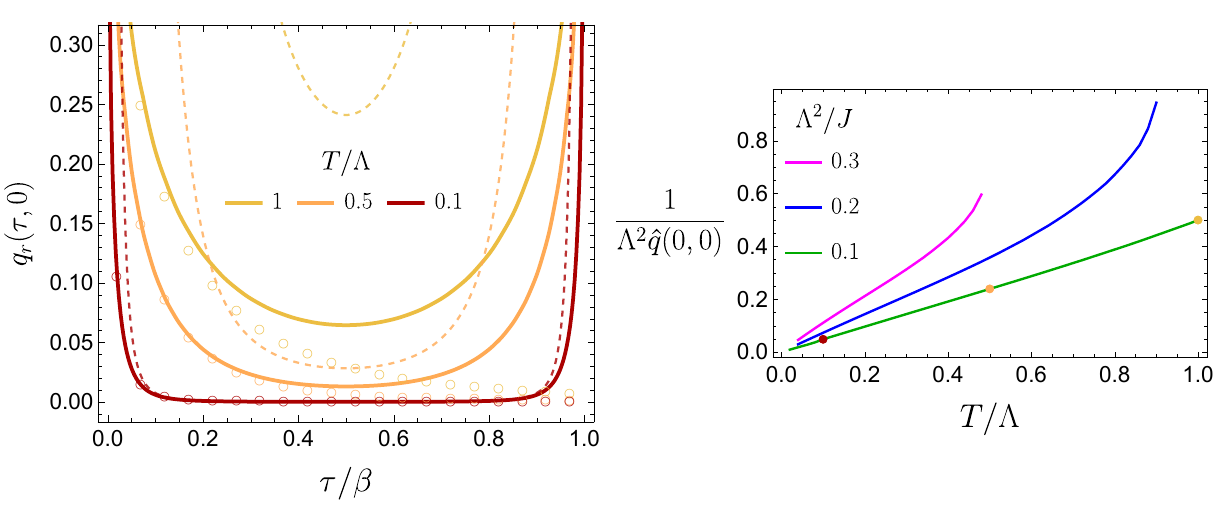}
    \caption{{\it Left}: Reduced correlator $q_r(\tau,0)$ in the deep spin glass regime, approaching the analytic approximation as $T\to 0$. Solid lines: full numerical solutions. Open circles: planar analytic approximation \eqref{eq:I0K0-I2K2}. Dashed lines: late-time conformal ansatz on the cylinder, Eq. \eqref{eq:conformalSGcylinder}. {\it Right}: Inverse zero modes as a function of the temperature for different values of the coupling $\Lambda^2/J$. They approach zero as $T\to 0$, consistent with the conformal-like scaling of the reduced correlator.}
    \label{fig:conformalSG}
\end{figure}

The central question is whether this phase realizes the scaling behavior anticipated by the analytic expectations of Section \ref{subsec:scaling}. To address this issue, we compare the numerical solutions with the analytic expressions. The comparison is shown in Fig. \ref{fig:conformalSG}.

The deep spin glass approximation derived analytically is quantitatively accurate. In the low-temperature regime, the numerical correlators are almost indistinguishable from the analytic approximation \eqref{eq:I0K0-I2K2}. Also in this limit, the late-time behavior is very well approximated by the scaling ansatz \eqref{eq:conformalSGcylinder}. This provides strong evidence that the analytical treatment captures the correct infrared dynamics of the glassy phase.

At the same time, the numerical solutions reveal an important limitation of the conformal picture. The scaling regime develops only after the Edwards-Anderson parameter $u$ has become non-zero. Consequently, the infrared behavior is never described by the replica-symmetric conformal solution of  \cite{Murugan:2017eto}. Instead, the scaling regime emerges within a phase that explicitly breaks replica symmetry.

Therefore, the role of the scaling solution is not to describe an actual conformal fixed point of the theory. Rather, the true infrared state is a spin glass exhibiting approximate scaling behavior rather than a conformal phase.

\subsection{Phase diagram and thermodynamics}

\begin{figure}
    \centering
    \adjustbox{valign=c}{
        \includegraphics[width=0.45\textwidth]{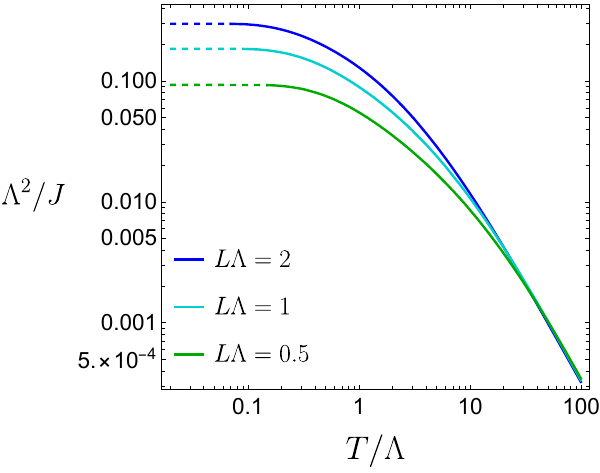}}
    \adjustbox{valign=c}{
        \includegraphics[width=0.48\textwidth]{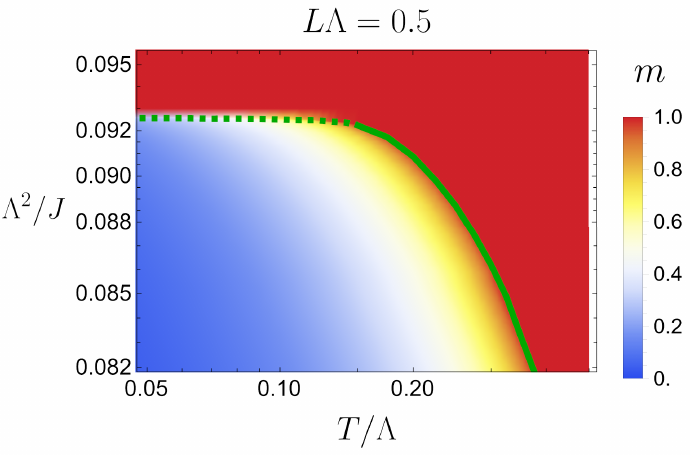}}
    \caption{{\it Left}: Phase diagram of the two-dimensional disordered $O(N)$ sigma model in the $(T/\Lambda,\Lambda^2/J)$ plane at fixed $L\Lambda$. The phase boundaries are determined by the onset of marginal spin-glass solutions. Dashed lines correspond to $T<T_c(L\Lambda)$. At these temperatures the transition is first order. Solid lines correspond to $T>T_c(L\Lambda)$, where the transition is second order. {\it Right}:  Order parameter $m$ across the first and second order transitions. The green lines correspond to the same green lines in the left plot.}
    \label{fig:phasediag}
\end{figure}

Having established the properties of the individual solutions, we now determine the complete phase diagram, which we show in Fig. \ref{fig:phasediag}. The overall structure is in striking agreement with the static approximation, despite the fact that the latter neglects all spatial and temporal fluctuations. For
temperatures above $T_c$, the transition between the paramagnetic and spin-glass phases is continuous: as $J$ is increased at fixed $T$ and $L$, a marginal spin-glass solution first appears at a critical coupling $J_c$, and the order parameter $m$ changes continuously from the paramagnetic value $m=1$. For $T < T_c$, the transition is first order: the system undergoes a direct PM$_g \to$ SG jump at $J = J_c$, where the SG solution begins to exist. We emphasize that the criterion for the onset of the spin-glass phase is the emergence of
a marginally stable solution to the equations~\eqref{eq:EoMqr}-\eqref{eq:marginalm}, not
the crossing of free energies.

The transition lines in the full theory are shifted quantitatively relative to the static approximation, but the main features of the phase diagram (including the existence of the tricritical point $T_c$ and the exchange of transition order across it) are preserved exactly. The behavior of the order parameter $m$ in the vicinity of the PM-to-SG transition for both $T < T_c$ and $T > T_c$ is illustrated in the right plot of Fig. \ref{fig:phasediag}.

We conclude by examining the thermodynamic observables obtained from the numerical solutions. In particular, we study the free energy, internal energy and specific heat for the two phases of the model using the thermodynamic expressions derived in Section \ref{sec:disordered}.

\begin{figure}[h!]
    \centering
    \includegraphics[width=0.48\textwidth]{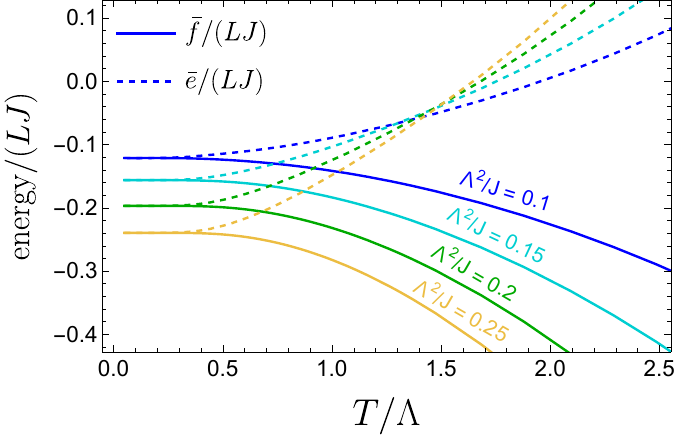}
    \includegraphics[width=0.48\textwidth]{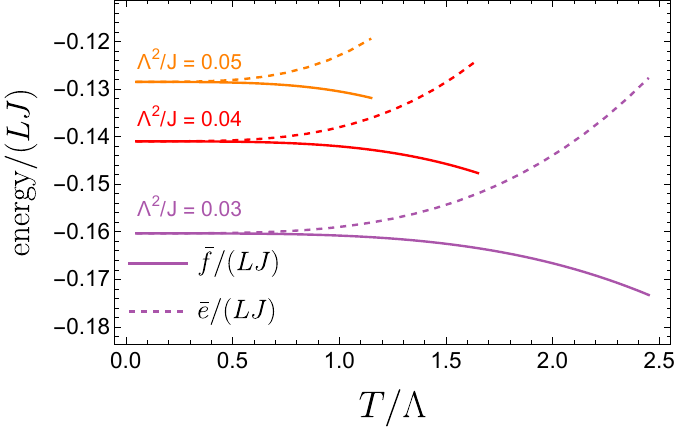}
    \caption{{\it Left}: Low-temperature behavior of the free energy (solid lines) and internal energy (dashed lines) for paramagnetic solutions with $L\Lambda=0.5$. At these values of $\Lambda^2/J$ the paramagnetic phase extends up to arbitrarily small temperatures (see Fig. \ref{fig:phasediag}). {\it Right}: Low-temperature behavior of the free energy (solid lines) and internal energy (dashed lines) for spin glass solutions, again for $L\Lambda=0.5$ but for lower values of $\Lambda^2/J$. At these values of the coupling the low-temperature regime is dominated by the spin glass phase.}
    \label{fig:FandE}
\end{figure}

It is interesting to study the behavior of the free energy as a function of the temperature once the coupling $\Lambda^2/J$ and the radius $L\Lambda$ are kept fixed. In Fig. \ref{fig:FandE} we consider a fixed value $L\Lambda=0.5$, and we show the free energy and the internal energy as a function of $T$. The curves are separated into two plots to distinguish between two conceptually different behaviors: on the left plot, we show the free energy and internal energy for relatively large values of $\Lambda^2/J$. Looking at the phase diagram of Fig. \ref{fig:phasediag}, for any fixed coupling $L\Lambda$, at sufficiently large $\Lambda^2/J$ the paramagnetic phase extends down to arbitrarily small temperatures. On the right plot, we show again the free energy and the internal energy, but for smaller values of $\Lambda^2/J$. These are solutions that at low temperatures transition into the spin-glass phase, as it can be seen again from the phase diagram in Fig. \ref{fig:phasediag}. 

We note that both $\Bar{e}/J$ and $\Bar{f}/J$ are monotonic and have vanishing slope at zero temperature. This is an indication that the 1-RSB ansatz considered here is thermodynamically consistent \cite{Sherrington:1975zz,mezard1987spin}.

Overall, the numerical results provide a detailed description of the infrared dynamics of the disordered sigma model. The qualitative phase structure anticipated from the static approximation is confirmed, and the analytical predictions for the deep spin glass regime are quantitatively reproduced. Most importantly, the numerics show that the infrared scaling behavior develops only inside the spin glass phase and not within the replica-symmetric phase. The theory therefore approaches a glassy scaling regime rather than a genuinely conformal infrared fixed point.

\section{Conclusions}

In this work, we studied a two-dimensional disordered QFT with bosonic degrees of freedom stabilized by a nonlinear sigma-model (spherical) constraint. We found that many of its properties, such as the phase structure, thermodynamics, and a low-temperature spin glass phase, are qualitatively similar to the quantum mechanical $p$-spin spherical model. The phase structure at low temperatures is intricate: in the paramagnetic phase, different solution branches exist, but a reparametrization invariant solution — suggested by a naive analysis of the Schwinger-Dyson equations — is avoided by the presence of a gap. On the other hand, the spin glass phase is characterized by a scaling solution, which is offset by the Edwards-Anderson order parameter.

A striking feature of the model considered here is the close similarity between the phase diagram of the present two-dimensional model and that of its one-dimensional counterpart studied
in~\cite{Cugliandolo_2001,Anous:2021eqj}. The organization of saddle-point solutions, their stability properties, and the first-order versus second-order structure of the transition lines are all qualitatively identical, despite the different ultraviolet structures of the one- and two-dimensional theories. This robustness with respect to dimensionality suggests that the physics of the glassy phase is governed by a universal mechanism that is independent of whether the model lives on a quantum-mechanical time line or on a two-dimensional torus.

It would be interesting to understand the scaling solution in the spin glass phase better and connect it to the framework of two-dimensional conformal field theory.\footnote{Some related work on glassy physics in QFT can be found in \cite{Sachdev_1995,Read_1995,da_Concei_o_2008,da_Concei_o_2009,daConceicao:2009zz}.} Relatedly, it would be exciting to develop a holographic framework for this phenomenon. In the context of AdS${}_2$/nCFT${}_1$ holography, a relation to AdS${}_2$ fragmentation \cite{Maldacena:1998uz} has been suggested \cite{Anous:2021eqj}. Our work suggests a similar picture in AdS${}_3$/CFT${}_2$, perhaps building on work such as \cite{deBoer:2008fk, Denef:2007vg,Denef:2011ee}. In the context of developing a holographic description involving multicentered black holes, it would also be interesting to compute real-time correlation functions and benchmark chaos in our model, for example by computing the quantum Lyapunov exponent in the glassy scaling regime (cf.\ \cite{Anous:2021eqj,Bera:2021lnh}) and the spectral form factor (cf.\ \cite{Winer:2022ciz}).

It also remains an open question whether a different stabilizing mechanism — other than the spherical constraint — could allow the bosonic MSW model to flow to its conformal infrared point without triggering a transition to a spin glass phase. Supersymmetric completions are expected to avoid the spin glass phase (see \cite{BerkoozToAppear}); whether a purely bosonic cure exists is still an open question. Relatedly, it would be interesting to study finite-$N$ effects and to simulate our model for individual realizations of the disorder, to investigate if our conclusions about the phase structure persist.

\acknowledgments

We thank Jessica Yeh for initial collaboration. We are also grateful to Tarek Anous, Jan Boruch, Gabriele Di Ubaldo, Adam Levine, Geoff Penington, and Moshe Rozali for useful discussions. FMH is  supported by UK Research and Innovation (UKRI) under the UK government’s Horizon Europe Funding Guarantee EP/X030334/1. The work of MB has been funded by Xunta de Galicia through the Programa de axudas \'a etapa predoutoral and the grant 2023-PG083 with reference code ED431F 2023/19.

\appendix
\section{Derivation of the large-\texorpdfstring{$\bd{N}$}{N} effective action and equations of motion}\label{app:EoM}

In this appendix we derive the Schwinger-Dyson equations of motion for the disordered $O(N)$ sigma model introduced in \eqref{eq:action},
\begin{equation}
    S = \int_{\cal M}d^2\xv\,\Bigl[\frac{1}{2}(\partial_\mu\sigma_i)^2
    + \frac{\lambda}{2}\!\left(\sigma_i^2 - \frac{N}{g_0^2}\right)
    + \sum_{i_1 < \cdots < i_p} J_{i_1\cdots i_p}\,
      \sigma_{i_1}\cdots\sigma_{i_p}
  \Bigr]\, .
\label{eq:actionApp}
\end{equation}

We choose the space to be the flat torus ${\cal M}=\mathbb{T}^2$, with periodicities $\beta$ and $L$. We follow the standard replica-field theory approach, integrating the quenched disorder to obtain a translationally-invariant effective action, and then imposing the one-step replica-symmetry-breaking (1-RSB) ansatz to arrive at the final momentum-space equations \eqref{eq:EoMqr}-\eqref{eq:renormalized constraint}.

Due to the possibility of a spin-glass transition, we "clone" the system $m$ times and introduce a small bias term in the action:
\begin{equation}
\label{eq:SclonedApp}
\begin{aligned}
    S_m &= \sum_{a=1}^m \intx\left[\frac{1}{2}(\partial_\mu\sigma^a_i)^2+\frac{\lambda^a}{2}\left( (\sigma_i^a)^2-\frac{N}{g_0^2}\right) +\sum_{i_1 < \cdots < i_p}J_{i_1 \cdots i_p}\sigma^a_{i_1}\cdots \sigma^a_{i_p}\right] \\
    &\quad + \epsilon\sum_{a,b=1}^m \int d^2\bd{x} \; \sigma^a_i\sigma^b_i \,.
\end{aligned}
\end{equation}

The disorder-averaged free energy in this cloned, grand-canonical ensemble is
\begin{equation}
\label{eq:PhiDefApp}
    \beta m \overline{\Phi} \equiv -\overline{\log Z[J_{i_1\cdots i_p}]^m} 
    =  - \lim_{n\rightarrow 0} \, \partial_n \, \overline{Z[J_{i_1\cdots i_p}]^{n m}} \,,
\end{equation}
where $Z[J_{i_1\cdots i_p}]$ is the partition function and the overline denotes a disorder average, defined as
\begin{equation}
\label{eq:defdisorder}
    \overline{Z[J_{i_1\cdots i_p}]} = \int {\cal D} J_{i_1\cdots i_p} \, P[J_{i_1\cdots i_p}] \, Z[J_{i_1\cdots i_p}] \,,\qquad 
    P\left[J_{i_1 \cdots i_p}\right] \propto \exp\left(-\frac{N^{p-1}}{p!}\frac{J^2_{i_1\cdots i_p}}{J^2}\right)\,.
\end{equation}
In addition to the original $m$ copies, we replicate the entire cloned system $n$ times to evaluate the logarithm; the second equality in \eqref{eq:PhiDefApp} is the standard replica trick for quenched free energies. We extend the indices $a,b=1,\dots,nm$ to label the combined replica space. The replicated partition function, after taking the $\epsilon\to 0$ limit, is
\begin{equation}
\begin{aligned}
    Z[J_{i_1\cdots i_p}]^{nm} = \int \mathcal{D}\sigma_i^a\,\mathcal{D}\lambda^a \exp\!&\left\{-\sum_{a=1}^{nm}\intx
    \left[\frac{1}{2}(\partial_\mu\sigma_i^a)^2+\frac{1}{2}\lambda^a\!\left((\sigma_i^a)^2 -\frac{N}{g_0^2}\right)\right. \right.\\&\left. \left.\phantom{\{}+ \sum_{i_1<\cdots<i_p} J_{i_1\ldots i_p}\,\sigma_{i_1}^a\cdots\sigma_{i_p}^a\right]\right\}\,.
\end{aligned}
\end{equation}
We now average $Z[J_{i_1\cdots i_p}]^{nm}$ over the disorder distribution according to \eqref{eq:defdisorder}. The disorder-dependent terms factorize into a product of independent Gaussian integrals that can be performed explicitly, which yields
\begin{equation}\label{eq:int_out_J}
\begin{aligned}
    \overline{Z^{nm}}=\int \mathcal{D}\sigma_i^a\,\mathcal{D}\lambda^a \exp\!&\left\{-\sum_{a}\intx
    \left[\frac{1}{2}(\partial_\mu\sigma_i^a)^2+\frac{1}{2}\lambda^a\!\left((\sigma_i^a)^2 -\frac{N}{g_0^2}\right)\right] \right.\\&\left. \phantom{\{}+ \frac{J^2}{4N^{p-1}}\sum_{a,b}\intxx\!\left(\sum_{i=1}^{N}\sigma_i^a(\xv)\,\sigma_i^b(\xvp)\right)^{\!p}\right\}\,,
\end{aligned}
\end{equation}
where we have made the replacement $p!\sum_{i_1<\cdots<i_p}\to\sum_{i_1,\cdots,i_p}$, which is valid in the large-$N$ limit.

We now introduce the mean-field (Hubbard-Stratonovich) variables $Q_{ab}(\xv,\xvp)$ and $\Sigma_{ab}(\xv,\xvp)$, where
\begin{equation}\label{eq:defQabApp}
    Q_{ab}(\xv,\xvp) \equiv \frac1N \sum_{i=1}^N \sigma_i^a(\xv) \sigma_i^b(\xvp)\,,
\end{equation}
by inserting the following factor into the path integral \eqref{eq:int_out_J}
\begin{equation}\label{eq:insert1}
\begin{aligned}
    1 & =\int \mathcal{D} Q_{a b}\, \prod_{a,b}\,\delta\!\left(NQ_{ab}(\xv,\xvp) -\sum_{i=1}^N \sigma_i^a(\xv) \sigma_i^b(\xvp)\right) \\
    & =\int \mathcal{D} Q_{a b} \mathcal{D} \Sigma_{ab} \exp \left\{i \sum_{a,b}\intxx \Sigma_{ab}(\xv,\xvp)\!\left(N Q_{ab}(\xv,\xvp)-\sum_{i=1}^{N}\sigma_i^a(\xv)\,\sigma_i^b(\xvp)\right)\right\}\,.
\end{aligned}
\end{equation}
This leads to
\begin{equation}
\begin{aligned}
    \overline{Z^{nm}}=\int& \mathcal{D}Q_{ab}\,\mathcal{D}\Sigma_{ab}\,\mathcal{D}\sigma_i^a\,\mathcal{D}\lambda^a\exp\left\{-\frac{1}{2}\sum_{a,b}\intxx
    \sigma_i^a(\xv)
    \left[
      \delta_{ab}\,\delta^{(2)}(\xv-\xvp)
      \left(-\partial_\mu^2 + \lambda^a(\xv)\right)\right.\right.\\
      &+ 2i\,\Sigma_{ab}(\xv,\xvp)
    \Bigr]\sigma_i^b(\xvp)+ \frac{N}{2g_0^2}\sum_a\intx\lambda^a(\xv)\\
    &+ N\sum_{a,b}\intxx\Bigl[
        i\,\Sigma_{ab}(\xv,\xvp)\,Q_{ab}(\xv,\xvp)
        + \frac{J^2}{4}Q_{ab}(\xv,\xvp)^p
      \Bigr]
  \Biggr\}\,.
\end{aligned}
\end{equation}

Since the action is now quadratic in the fields $\sigma_i^{a}$, the $N$ identical Gaussian integrals can be performed exactly, leading to
\begin{equation}
\begin{aligned}
    \overline{Z^{nm}}&=\int \mathcal{D}Q_{ab}\,\mathcal{D}\Sigma_{ab}\,\mathcal{D}\lambda^a\det\left[\frac{1}{2\pi}\delta_{ab}\,\delta^{(2)}(\xv-\xvp)\!\left(-\partial_\mu^2 + \lambda^a(\xv)\right)+\frac{i}{\pi}\Sigma_{ab}(\xv,\xvp)\right]^{-N/2} \\ &\times\exp\Biggl\{ \frac{N}{2g_0^2}\sum_a\intx\lambda^a(\xv)+ N\sum_{a,b}\intxx\Bigl[
        i\,\Sigma_{ab}(\xv,\xvp)\,Q_{ab}(\xv,\xvp)
        + \frac{J^2}{4}Q_{ab}(\xv,\xvp)^p
      \Bigr]
  \Biggr\}\,,
\end{aligned}
\end{equation}
which can be written as
\begin{equation}\label{eq:ZavSeff}
    \overline{Z^{nm}}=\int\mathcal{D}Q_{ab}\,\mathcal{D}\Sigma_{ab}\,\mathcal{D}\lambda^ae^{-\Seff[Q,\Sigma,\lambda]}\,,
\end{equation}
with the large-$N$ effective action
\begin{equation}\label{eq:SefflargeNApp}
\begin{aligned}
    \frac{\Seff}{N}
  &= \frac{1}{2}\Tr\log\left[\frac{1}{2\pi}\delta_{ab}\,\delta^{(2)}(\xv-\xvp)\!\left(-\partial_\mu^2 + \lambda^a(\xv)\right)+\frac{i}{\pi}\Sigma_{ab}(\xv,\xvp)\right]\\
  &\quad- \frac{1}{2g_0^2}\sum_a\intx\lambda^a(\xv)-\sum_{a,b}\intxx\!\left[i\,\Sigma_{ab}(\xv,\xvp)\,Q_{ab}(\xv,\xvp)+\frac{J^2}{4}Q_{ab}(\xv,\xvp)^p\right]\,.
\end{aligned}
\end{equation}
Because $\Seff\propto N$, the saddle-point approximation becomes exact in the large-$N$ limit, and the integral \eqref{eq:ZavSeff} is dominated by the stationary points of $\Seff$.

Varying $\Seff$ with respect to $\Sigma_{ab}$ we obtain the first equation of motion,
\begin{equation}\label{eq:EoMSigmaab}
    (\bd{\mathrm{M}}+i\bd{\Sigma})^{-1}=2\bd{\mathrm{Q}}\,,
\end{equation}
where the matrix $\bd{\mathrm{M}}$ is defined as
\begin{equation}
  M_{ab}(\xv,\xvp)\equiv\frac{1}{2}\delta_{ab}\,\delta^{(2)}(\xv-\xvp)\!\left(-\partial_\mu^2 + \lambda^a(\xv)\right)\,.
\label{eq:M_def}
\end{equation}
Substituting \eqref{eq:EoMSigmaab} back into the effective action \eqref{eq:SefflargeNApp}, we find
\begin{equation}\label{eq:SeffQabApp}
\begin{aligned}
    \frac{\Seff}{N}&=-\frac{1}{2}\Tr\log Q_{ab}+\frac{1}{2}\sum_a\intx\lambda^a(\xv)\!\left(Q_{aa}(\xv,\xv) - \frac{1}{g_0^2}\right)\\
    &\quad-\sum_{a,b}\intxx\!\left[\delta_{ab}\,\delta^{(2)}(\xv-\xvp)\,\frac{1}{2}\partial_\mu^2\,Q_{ab}(\xv,\xvp)+\frac{J^2}{4}\,Q_{ab}(\xv,\xvp)^p\right]\,
\end{aligned}
\end{equation}
up to irrelevant constant terms.

We can obtain a single equation of motion for $Q_{ab}$ varying \eqref{eq:SeffQabApp},
\begin{equation} \label{eq:Schwinger-Dyson}
    -\delta_{ab}\left[\partial_\mu^2 - \lambda^a(\xv)\right]Q_{ab}(\xv,\xvp) - \frac{p J^2}{2}\sum_c\intxpp\, Q_{ac}^{p-1}(\xv,\xvpp)Q_{cb}(\xvpp,\xvp) = \delta_{ab}\,\delta^{(2)}(\xv-\xvp)\,.
\end{equation}

We now specialize to the 1-RSB ansatz. The $nm\times nm$ matrix $Q_{ab}$ is taken to have a block structure with blocks of size $m$: within each block the diagonal entries are equal and given by the connected two-point function $q(\xv,\xvp)$, and the off-diagonal entries equal the Edwards-Anderson parameter $u$. All inter-block entries vanish:
\begin{equation}
\label{eq:1rsbApp}
	Q_{ab}(\xv,\xvp) =\begin{pmatrix} 
	q(\bd{x},\bd{x}') & u & u &   &  & &\\
	u & q(\bd{x},\bd{x}') &u &  & 0& &\cdots\\
	u & u & q(\bd{x},\bd{x}') & & & &\\
	  &  &  & q(\bd{x},\bd{x}') &  u& u &\\
	  &  0&  & u & q(\bd{x},\bd{x}') &u &\\
	  &  &  & u & u & q(\bd{x},\bd{x}') &\\
	  & \vdots & & & & & \ddots &
	\end{pmatrix} \,.
\end{equation}
The matrix \eqref{eq:1rsbApp} has the following eigenvalues:
\begin{equation}
\begin{aligned}
    \lambda_1 &=q(\xv,\xvp)-u\,,&\qquad &\text{degeneracy}~nm\left(1-\frac{1}{m}\right)\,,\\
    \lambda_2 &=q(\xv,\xvp)+u(m-1)\,,&\qquad &\text{degeneracy}~n\,.
\end{aligned}
\end{equation}
It is also useful to count the number of times $q(\xv,\xvp)$ and $u$ appear in \eqref{eq:1rsbApp}. It is immediate to see that
\begin{equation}
    \#\,q(\xv,\xvp)=nm\,,\qquad \#\,u=nm(m-1)\,.
\end{equation}

Using all these results, we can evaluate the effective action \eqref{eq:SeffQabApp} for the 1-RSB ansatz:
\begin{equation}\label{eq:Seffqum}
\begin{aligned}
    \frac{\Seff}{nmN}&=-\frac{1}{2}\!\left(1-\frac{1}{m}\right)\Tr\log\!\left(q(\xv,\xvp)-u\right)-\frac{1}{2m}\Tr\log\!\left(q(\xv,\xvp)+u(m-1)\right)\\
    &\quad+ \frac{1}{2}\intx\lambda(\xv)\!\left(q(\xv,\xv) -\frac{1}{g_0^2}\right)\\
    &\quad- \frac{1}{2}\intxx\!\left[\delta^{(2)}(\xv-\xvp)\,\partial_\mu^2\,q(\xv,\xvp)+ \frac{J^2}{2}\!\left(q(\xv,\xvp)^p + (m-1)\,u^p\right)\right]\,,
\end{aligned}
\end{equation}
where we have also assumed that $\lambda^a$ is independent of the replica index, $\lambda^a=\lambda$.

Finally, it will be convenient to write the effective action and the equations of motion in momentum space. Our convention for the Fourier transforms is
\begin{equation}
    q(\xv) = \frac{1}{\beta L}\sum_{k,l} e^{-i(\omega_k\tau + p_l x)}\,\hq(k,l)\,,\qquad\hq(k,l) = \intx\, e^{i(\omega_k\tau+p_l x)}\,q(\xv)\,,
\label{eq:fourier}
\end{equation}
where the Matsubara frequencies and momenta are given by $\omega_k=2\pi k/\beta$ and $p_l=2\pi l/L$, respectively, and $\xv=(\tau,x)$.

The effective action \eqref{eq:Seffqum} becomes
\begin{equation}\label{eq:Seffqhatum}
\begin{aligned}
    \frac{\Seff}{nmN}&=  -\frac{1}{2}\sum_{(k,l)\neq(0,0)}\log\!\frac{\hq(k,l)}{\beta L}-\frac{1}{2}\log\!\left[\frac{\hq(0,0)}{\beta L}-u\right]+\frac{1}{2m}\log\left[\frac{\frac{\hq(0,0)}{\beta L}-u}{\frac{\hq(0,0)}{\beta L}+u(m-1)}\right]\\
    &\quad+\frac{1}{2}\,\hat\lambda(0,0)\left(\sum_{k,l}\frac{\hq(k,l)}{\beta L} - \frac{1}{g_0^2}\right)+\frac{1}{2}\sum_{k,l}\!\left[\left(\frac{2\pi k}{\beta}\right)^2+\left(\frac{2\pi l}{L}\right)^2\right]\hq(k,l)\\
    &\quad-\frac{(\beta L J)^2}{4}\!\Biggl[(m-1)\,u^p+ \sum_{\substack{k_1,\ldots,k_p\\l_1,\ldots,l_p}}\prod_{i=1}^p\!\left[\frac{\hq(k_i,l_i)}{\beta L}\right]\delta_{k_1+\cdots+k_p,\,0}\;\delta_{l_1+\cdots+l_p,\,0}\Biggr]\,.
\end{aligned}
\end{equation}

Before proceeding, it is convenient to define the self-energy $\Lambda(\xv,\xvp)$ as
\begin{equation}
    \Lambda(\xv,\xvp)\equiv\frac{p}{2}q(\xv,\xvp)^{p-1}\,,
\end{equation}
whose Fourier transform is
\begin{equation} 
    \hat{\Lambda}(k,l) = \frac{p}{2(\beta L)^{p-2}} \sum_{\substack{k_1,\ldots,k_{p-2}\\l_1,\ldots,l_{p-2}}} \hat{q}(k_1,l_1)\cdots\hq(k_{p-2},l_{p-2})\hq(k-k_1-\ldots-k_{p-2},l-l_1-\ldots-l_{p-2})\,.
\end{equation}
Varying \eqref{eq:Seffqhatum} with respect to $\hq(k,l)$ gives the equation of motion in momentum space:
\begin{equation}\label{eq:Eomqhat}
\begin{split}
&\left(\frac{2\pi k}{\beta}\right)^2+\left(\frac{2\pi l}{L}\right)^2+\frac{\hat\lambda(0,0)}{\beta L} \\
&\qquad =\frac{\hq(k,l)+\beta L\,u(m-2)\,\delta_{k,0}\delta_{l,0}}{\hq(k,l)^2-\beta L\,u\,\delta_{k,0}\delta_{l,0}\bigl[(m-1)\beta L\,u-(m-2)\hq(k,l)\bigr]}+J^2\hL(k,l)\,.
\end{split}
\end{equation}
We can similarly vary with respect to $u$ to obtain
\begin{equation}\label{eq:Eomu}
    \frac{m-1}{2}\!\left[\frac{(\beta L J)^2}{2}\,p\,u^{p-1}-\frac{u}{\!\left(\frac{\hq(0,0)}{\beta L}-u\right)\!\left(\frac{\hq(0,0)}{\beta L}+u(m-1)\right)}\right]=0\,.
\end{equation}
Varying with respect to $\hat{\lambda}(0,0)$ we recover the spherical constraint
\begin{equation}
    \sum_{k,l}\frac{\hq(k,l)}{\beta L} = \frac{1}{g_0^2}\,.
\end{equation}

Finally, we can simplify the momentum space equations \eqref{eq:Eomqhat}-\eqref{eq:Eomu} if we define the "reduced" variables
\begin{equation}
    q_r(\xv,\xvp)\equiv q(\xv,\xvp)-u\,,\qquad \Lambda_r(\xv,\xvp) \equiv\Lambda(\xv,\xvp)-\frac{p}{2}u^{p-1}\,,
\end{equation}
which in momentum space translate into
\begin{equation}
    \hq_r(k,l)= \hq(k,l)-\beta L u\,\delta_{k,0}\,\delta_{l,0}\,,\qquad \hL_r(k,l)=\hL(k,l)-\frac{p}{2}\beta L u^{p-1}\delta_{k,0}\delta_{l,0}\,.
\end{equation}
With these definitions, the first term on the rhs of \eqref{eq:Eomqhat} simplifies significantly and we obtain
\begin{equation}
    \left(\frac{2\pi k}{\beta}\right)^2+ \left(\frac{2\pi l}{L}\right)^2+\frac{\hat\lambda(0,0)}{\beta L}=\frac{1}{\hq_r(k,l)}+J^2\hL_r(k,l)\,.
\end{equation}
Evaluating this last expression for $(k,l)=(0,0)$ allows to solve for the Lagrange multiplier:
\begin{equation}
    \frac{\hat{\lambda}(0,0)}{\beta L}=\frac{1}{\hq_r(0,0)}+J^2\hL_r(0,0)\,,
\end{equation}
allowing us to write the final equations of motion, 
\begin{align}
    &\text{EoM}_{\hq_r}: &\frac{1}{\hq_r(k,l)} - \frac{1}{\hq_r(0,0)}&=\left(\frac{2\pi k}{\beta}\right)^2+\left(\frac{2\pi l}{L}\right)^2-J^2\!\left(\hL_r(k,l) - \hL_r(0,0)\right)\,,\\
    &\text{EoM}_{u}: &0&=\frac{m-1}{2}\left[\frac{(\beta LJ)^2}{2}p\,u^{p-1}-\frac{u}{\frac{\hq_r(0,0)}{\beta L}\left(\frac{\hq_r(0,0)}{\beta L}+mu\right)}\right]\,.
\end{align}

Recall that these equations are supplemented by the extra condition on the parameter $m$ obtained by imposing marginal stability, cf.\ \eqref{eq:marginalm},
\begin{equation}
    \text{EoM}_{m}: \qquad m = \frac{p-2}{\beta L \mathcal{J} \, u^{p/2}}\,,\qquad \mathcal{J}\equiv J\sqrt{\frac{p(p-1)}{2}}\,.
\end{equation}

\section{Regularization of the free energy}\label{app:FreeEnReg}

In this appendix we regularize the divergences of the on-shell effective action \eqref{eq:Seff} to obtain the different thermodynamic quantities.

Evaluating \eqref{eq:Seff} on shell, the term imposing the spherical constraint vanishes,
\begin{equation}
\begin{aligned}
   \beta  \overline{\phi}&= \frac{S_{\rm eff}}{nmN}\\
   &= -\frac{1}{2}\sum_{(k,l)\neq(0,0)}\log\!\frac{\hq(k,l)}{\beta L}-\frac{1}{2}\!\left(1-\frac{1}{m}\right)\log\!\left[\frac{\hq(0,0)}{\beta L}-u\right]-\frac{1}{2m}\log\!\left[\frac{\hq(0,0)}{\beta L}+u(m-1)\right]\\
    &\quad+\frac{1}{2}\sum_{k,l}\!\left[\left(\frac{2\pi k}{\beta}\right)^2+\left(\frac{2\pi l}{L}\right)^2\right]\hq(k,l)-\frac{(\beta L J)^2}{4}\!\Biggl[(m-1)\,u^p+ \frac{1}{\beta L}\intx\,q(\bd{x})^p\Biggr]\,.
    \label{eq:Seffonshell}
\end{aligned}
\end{equation}
The two infinite sums appearing in this expression are naively UV divergent. To regularize them, we introduce the "UV solution"
\begin{equation}
    \frac{\hat{q}_r^{\rm UV}(k,l)}{\beta L} = \frac{1}{\beta L}\left[\left(\frac{2\pi k}{\beta}\right)^2+\left(\frac{2\pi l}{L}\right)^2+\lambda \right]^{-1}\,,
\end{equation}
where $\lambda=\frac{\hat{\lambda}(0,0)}{\beta L}$ is the (constant) position space Lagrange multiplier. By adding and subtracting $\Seff$ evaluated on the UV solution, we obtain
\begin{equation}
\begin{aligned}
    \frac{S_{\rm eff}}{nmN} =&-\frac{1}{2}\sum_{k,l}\log \left\{ \left[\left(\frac{2\pi k}{\beta}\right)^2 + \left(\frac{2\pi l}{L}\right)^2+\lambda\right]\hat{q}_r(k,l)\right\} + \frac{1}{2m}\log \left[\frac{\frac{\hat{q}_r(0,0)}{\beta L}}{\frac{\hat{q}_r(0,0)}{\beta L}+ m u}  \right]\\
    & -\frac{1}{2} \sum_{k,l} \left\{1-\left[\left(\frac{2\pi k}{\beta}\right)^2 + \left(\frac{2\pi l}{L}\right)^2 +\lambda\right]\hat{q}_r(k,l) \right\} \\
    & - \frac{(\beta LJ)^2 }{4} \left[ (m-1)u^p + \frac{1}{\beta L} \int_\mathcal{M} d^2\bd{x}\ q(\bd{x})^p \right] -\frac{\beta L \lambda}{2}\sum_{k,l} \frac{\hat{q}_r(k,l)}{\beta L}\\
    &+ \frac{1}{2} \sum_{k,l}  \left\{ 1 +\log(\beta L)+ \log\left[\left(\frac{2\pi k}{\beta}\right)^2 + \left(\frac{2\pi l}{L}\right)^2+\lambda\right] \right\}  \,.
    \label{eq:Seffaddsubtr}
\end{aligned}
\end{equation}
The term proportional to $\beta L \lambda$ evaluates to
\begin{equation}
    -\frac{\beta L \lambda}{2}\sum_{k,l} \frac{\hat{q}_r(k,l)}{\beta L}=-\frac{\beta L \lambda}{2}\left(\frac{1}{g_0^2}-u\right)~,
\end{equation}
where the bare ultraviolet coupling $1/g_0^2$ can be absorbed into a UV bare energy density, following the prescription of \cite{Bolognesi:2019rwq}. The remaining finite contribution is $\frac{\beta L\lambda\,u}{2}$.

The sums appearing in the first two lines of \eqref{eq:Seffaddsubtr} are now convergent at large momenta. The remaining divergent contribution arises from the last term. Using $\zeta$-function regularization, the $k,l$-independent term $\sum_{k,l}\left(1+\log(\beta L)\right)$ vanishes, and we are left with the sum
\begin{equation}
    \frac{1}{2}\sum_{k,l}\log\left[\left(\frac{2\pi k}{\beta}\right)^2 + \left(\frac{2\pi l}{L}\right)^2+\lambda \right]\,.
    \label{eq:logtoreg}
\end{equation}
This expression coincides with the logarithmic determinant regularized via Pauli–Villars in Ref.~\cite{Bolognesi:2019rwq}. Proceeding analogously, one finds
\begin{equation}
\begin{aligned}
    \frac{1}{2}\sum_{k,l}&\log\left[\left(\frac{2\pi k}{\beta}\right)^2 + \left(\frac{2\pi l}{L}\right)^2+\lambda \right]\Bigg\rvert_{reg}\\
    &=-\beta L\left[\frac{\lambda}{8\pi}\log\frac{\lambda}{e\Lambda^2}+\frac{1}{2\pi}\sum_{(k,l)\neq (0,0)}\sqrt{\frac{\lambda}{\left((k\beta)^2+(lL)^2\right)}}K_1\left(\sqrt{\lambda ((k\beta)^2+(lL)^2)}\right)\right]~,
    \label{eq:sumlogsreg}
\end{aligned}
\end{equation}
where $K_1$ is the modified Bessel function of the second kind. Putting all contributions together, the regularized effective action reads
\begin{equation}
\begin{aligned}
    \frac{\Seff^{\rm reg}}{nmN}=&-\frac{1}{2}\sum_{k,l}\log \left\{ \left[\left(\frac{2\pi k}{\beta}\right)^2 + \left(\frac{2\pi l}{L}\right)^2+\lambda\right]\hat{q}_r(k,l)\right\} + \frac{1}{2m}\log \left[\frac{\frac{\hat{q}_r(0,0)}{\beta L}}{\frac{\hat{q}_r(0,0)}{\beta L}+ m u}  \right]\\
    & -\frac{1}{2} \sum_{k,l} \left\{1-\left[\left(\frac{2\pi k}{\beta}\right)^2 + \left(\frac{2\pi l}{L}\right)^2 +\lambda \right]\hat{q}_r(k,l) \right\}+\frac{\beta L\lambda \,u}{2} \\
    & - \frac{(\beta LJ)^2 }{4} \left[ (m-1)u^p + \frac{1}{\beta L} \int_\mathcal{M} d^2\bd{x}\ q(\bd{x})^p \right] \\
    &-\beta L\left[\frac{\lambda}{8\pi}\log\frac{\lambda}{e\Lambda^2}+\frac{1}{2\pi}\sum_{(k,l)\neq (0,0)}\sqrt{\frac{\lambda}{\left((k\beta)^2+(lL)^2\right)}}K_1\left(\sqrt{\lambda ((k\beta)^2+(lL)^2)}\right)\right]  \,.
\end{aligned}
\end{equation}

\section{Numerical implementation}\label{app:numericalmethod}

In this appendix we briefly summarize the numerical procedure used to solve the Schwinger-Dyson equation together with the replica equations determining the Edwards-Anderson parameter $u$ and the replica-symmetry-breaking parameter $m$. Our goal is not to provide an exhaustive discussion of numerical methods but rather to describe the main ingredients of the algorithm used to generate the results presented in Section \ref{sec:numerical}.

The numerical solutions are obtained by iteratively solving the equations \eqref{eq:EoMqr}-\eqref{eq:marginalm}. The algorithm is implemented in momentum space and makes extensive use of discrete Fourier transforms to evaluate the non-linear self-energy $\Lambda$.

The theory is defined on a Euclidean torus of temporal circumference $\beta$ and spatial circumference $L$. We discretize both directions by introducing Matsubara frequencies and spatial momenta given by
\begin{equation}
    \omega_k=\frac{2\pi k}{\beta}\,,\qquad p_l=\frac{2\pi l}{L}\,,
\end{equation}
respectively, subject to an ultraviolet cutoff $\Lambda_{\rm UV}$. In practice we retain all modes satisfying
\begin{equation}
    \omega_k^2+p_l^2\le \Lambda_{\rm UV}^2\,.
\end{equation}

The propagator $\hq_r(k,l)$ is therefore represented as a finite two-dimensional array in momentum space.

For the evaluation of the interaction term it is necessary to transform repeatedly between momentum and position space. To this end, we introduce uniform lattices in Euclidean time and space,
\begin{equation}
    \tau_i\in [0,\beta)\,,\qquad x_j\in [0,L)\,,
\end{equation}
and use discrete Fourier transforms to relate the two representations.

The numerical procedure starts from an initial guess for the propagator in momentum space, $\hq_r^{(0)}(k,l)$. For the first point in parameter space we use the "free" massive solution \eqref{eq:sphconstr}, together with $(u,m)=(0,1)$. For subsequent points we employ the previously converged solution as the initial condition. This continuation procedure improves convergence and allows us to track the different branches of solutions when regions of coexistence emerge.

Given $\hq_r^{(i)}(k,l)$ and $u$ at iteration step $i$, we Fourier transform to position space and compute the self-energy
\begin{equation}
    \Lambda_r^{(i)}(\xv)=\frac{p}{2}\left[\bigl(q_r^{(i)}(\xv)+u\bigr)^{p-1}-u^{p-1}\right]\,.
\end{equation}
The resulting self-energy is then transformed back to momentum space and inserted into the Schwinger-Dyson equation to obtain an updated propagator, $\hq_r^{\rm \,upd}(k,l)$.

To improve numerical stability we do not replace the old propagator by the new one directly. Instead, we use a weighted update scheme,
\begin{equation}
    \hq_r^{(i+1)}(k,l)=(1-x)\,\hq_r^{(i)}(k,l)+x\,\hq_r^{\rm\, upd}(k,l)\,,
\end{equation}
with $0<x<1$. Since the Schwinger-Dyson equation \eqref{eq:EoMqr} is empty for $(k,l)=(0,0)$, we use the spherical constraint \eqref{eq:renormalized constraint} to determine the zero mode $\hq_r(0,0)$ at each iteration.

An important feature of the algorithm is that the replica parameters $u$ and $m$ cannot be determined independently of the propagator. After each Schwinger-Dyson update, the value of the zero mode is inserted into the replica equations. Depending on the phase under consideration, different equations are used: for the paramagnetic solutions, we simply set $(u,m)=(0,1)$. For the spin glass solutions, $u$ and $m$ are obtained by satisfying simultaneously the equation of motion for $u$, Eq. \eqref{eq:usol}, and the condition of marginal stability for $m$, Eq. \eqref{eq:marginalm}.

The values of $u$ and $m$ obtained in this way are then fed back into the self-energy and used in the next Schwinger-Dyson iteration.

The iterative procedure is repeated until the equation of motion \eqref{eq:EoMqr} is satisfied with a prescribed tolerance. Typical solutions converge after a few tens or hundreds of iterations. Once convergence is achieved, the resulting propagator, the self-energy, the Edwards-Anderson parameter $u$ and the replica parameter $m$ are extracted and can be used to compute the different thermodynamic quantities presented in Section \ref{sec:numerical}.

\section{Dimensional reduction to 1D} \label{app:dim_red}

We now study the limit in which the spatial circle is the smallest scale in the system,
\begin{equation}
    L\ll \beta\,,\quad L\Lambda\ll1\,,
\end{equation}
for which the model is expected to reduce to the $(0+1)$-dimensional quantum $p$-spin model \cite{Anous:2021eqj}. In this regime, the non-zero spatial momentum modes, with $p_l=\frac{2\pi l}{L}$, are parametrically heavy and can be integrated out. The low-energy dynamics are therefore dominated by the spatial zero mode $p_l=0$.

It is useful to make the normalization of this zero mode explicit. We define
\begin{equation}
    M_{1/L}\equiv \frac{1}{g(\mu=1/L)^2} = \frac{1}{4\pi}\log\frac{1}{L^2\Lambda^2}\,,
\end{equation}
where the second equality follows from the dimensional transmutation discussed in Eq. \eqref{eq:Lambda}. The zero-mode field is then normalized as
\begin{equation}
    \sigma_i(\tau,x) = \sqrt{M_{1/L}}\,s_i(\tau)+\cdots \,,
\end{equation}
where the ellipsis denotes the massive spatial modes. With this normalization, the two-dimensional kinetic term reduces to
\begin{equation}
    \frac12\int_0^\beta d\tau \int_0^L dx\,(\partial_\tau\sigma_i)^2 = \frac{M_\text{1d}}{2} \int_0^\beta d\tau\,(\partial_\tau s_i)^2 \,, \qquad M_\text{1d} \equiv L M_{1/L} \,.
\end{equation}
Thus the one-dimensional ``mass'' parameter is not an independent remnant of the canonical two-dimensional kinetic term; it is the running inverse coupling evaluated at the compactification scale, multiplied by the length of the spatial circle. Similarly, the disorder interaction reduces to
\begin{equation}
    \int_0^\beta d\tau \int_0^L dx\,J^{(\text{2d})}_{i_1\cdots i_p}\,\sigma_{i_1}\cdots \sigma_{i_p}=\int_0^\beta d\tau\,J^{(\text{1d})}_{i_1\cdots i_p}\,s_{i_1}\cdots s_{i_p}\,,
\end{equation}
with
\begin{equation}
    J^{(\text{1d})}_{i_1\cdots i_p}=L\, M_{1/L}^{p/2} J^{(\text{2d})}_{i_1\cdots i_p}\,,\quad J_\text{1d}^2 = L^2 M_{1/L}^pJ_\text{2d}^2\,.
\end{equation}

The strict dimensional reduction to a non-trivial quantum-mechanical $p$-spin model is therefore a double-scaling limit:
\begin{equation}\label{eq:ds}
    L\ll \beta\,,\qquad M_{1/L}\gg1\,,\qquad L M_{1/L}\to M_\text{1d}\,,\qquad L^2 M_{1/L}^pJ_\text{2d}^2 \to J_\text{1d}^2\,.
\end{equation}
Keeping $M_\text{1d}$ finite requires $e^{-\frac{2\pi}{L}M_\text{1d}} = L\Lambda\ll1$, so the dynamically generated scale $\Lambda$ is much smaller than the compactification scale $1/L$, and hence the lower bound \eqref{eq:zmplanar} becomes irrelevant in this dimensional reduction.

Let us now show how the Schwinger-Dyson equations reduce. We use the Fourier conventions of Eq. \eqref{eq:fourier},
\begin{equation}
    q_\text{2d}(\tau,x) = \frac{1}{\beta L}\sum_{k,l}e^{-i(\omega_k\tau+p_l x)}\hat q_\text{2d}(k,l)\,,\qquad q_\text{1d}(\tau)=\frac{1}{\beta}\sum_k e^{-i\omega_k\tau}\hat q_\text{1d}(k)\,,
\end{equation}
where $\omega_k=2\pi k/\beta$ and $p_l=2\pi l/L$. We identify the one- and two-dimensional correlators via
\begin{equation}\label{eq:identif}
\begin{split}
    q_\text{2d}(\tau,x) = M_{1/L} q_\text{1d}(\tau)\,,\quad &u_\text{2d} = M_{1/L} u_\text{1d} \,, \\
    \hat q_\text{2d}(k,0)
    = M_{1/L}L\,\hat q_\text{1d}(k)\,,\quad &\hat q_{r,\text{2d}}(k,0)
    = M_{1/L}L\,\hat q_{r,\text{1d}}(k)\,.
\end{split}
\end{equation}
This identification also fixes a similar scaling for the self-energies,
\begin{equation}
    \hat{\Lambda}_{r,\text{2d}}(k,0) = M_{1/L}^{p-1}L\,\hat{\Lambda}_{r,\text{1d}}(k)\,.
\end{equation}
The renormalized spherical constraint \eqref{eq:renormalized constraint} 
\begin{equation}
    \frac{1}{g(\mu)^2} = \sum_{|\bd{k}|^2<\Lambda_{UV}^2}\frac{\hq(k,l)}{\beta L} - \frac{1}{4\pi}\log\frac{\Lambda_{\rm UV}^2}{\mu^2} \,,
\end{equation}
where $|\bd{k}|^2\equiv\omega_k^2+p_l^2$,
simplifies to
\begin{equation}
\begin{split}
    M_{1/L} &= \frac{1}{\beta L}\sum_k \hat q_\text{2d}(k,0) + \left[\frac{1}{\beta L}\sum_{\substack{|\bd{k}|^2<\Lambda_{UV}^2\\l\neq0}} \hat q_\text{2d}(k,l) - \frac{1}{4\pi} \log(\Lambda_{\rm UV}^2L^2)\right] \\
    &= \frac{M_{1/L}}{\beta}\sum_k \hat q_\text{1d}(k) + \O(1) \,.
\end{split}
\end{equation}
Hence, after dividing by $M_{1/L}$, we recover the one-dimensional spherical constraint
\begin{equation}
    1=\frac{1}{\beta}\sum_k \hat q_\text{1d}(k)
\end{equation}
in the strict dimensional reduction limit.

Keeping only the spatial zero mode in the two-dimensional Schwinger-Dyson equation \eqref{eq:EoMqr} gives
\begin{equation}
    \frac{1}{\hat{q}_{r,\text{2d}}(k,0)}-\frac{1}{\hat{q}_{r,\text{2d}}(0,0)} = \omega_k^2-J_\text{2d}^2\left(\hat{\Lambda}_{r,\text{2d}}(k,0)-\hat{\Lambda}_{r,\text{2d}}(0,0)\right)\,.
    \label{eq:SDzeromode}
\end{equation}
Multiplying by $M_{1/L} L$ and using the scaling relations \eqref{eq:ds} and \eqref{eq:identif}, this becomes
\begin{equation}
    \frac{1}{\hat{q}_{r,\text{1d}}(k)}-\frac{1}{\hat{q}_{r,\text{1d}}(0)}=M_\text{1d}\omega_k^2-J_\text{1d}^2\left(\hat{\Lambda}_{r,\text{1d}}(k)-\hat{\Lambda}_{r,\text{1d}}(0)\right)\,,
\end{equation}
which, together with the spherical constraint, reproduces the Schwinger-Dyson equations of motion for the one-dimensional quantum spherical $p$-spin model of \cite{Anous:2021eqj}. The equations of motion for $u$ and $m$ can be derived from the two-dimensional model in an analogous manner.

\bibliographystyle{JHEP}
\bibliography{ref}

\clearpage

\end{document}